\documentclass[twocolumn,english,aps,prstab,groupedaddress]{revtex4}
\usepackage[latin9]{inputenc}
\usepackage{xcolor}
\usepackage{pdfcolmk}
\usepackage{units}
\usepackage{textcomp}
\usepackage{amsmath}
\usepackage{amssymb}
\usepackage{graphicx}
\PassOptionsToPackage{normalem}{ulem}
\usepackage{ulem}

\makeatletter

\providecommand{\tabularnewline}{\\}
\providecolor{lyxadded}{rgb}{0,0,1}
\providecolor{lyxdeleted}{rgb}{1,0,0}

\@ifundefined{textcolor}{}
{%
 \definecolor{BLACK}{gray}{0}
 \definecolor{WHITE}{gray}{1}
 \definecolor{RED}{rgb}{1,0,0}
 \definecolor{GREEN}{rgb}{0,1,0}
 \definecolor{BLUE}{rgb}{0,0,1}
 \definecolor{CYAN}{cmyk}{1,0,0,0}
 \definecolor{MAGENTA}{cmyk}{0,1,0,0}
 \definecolor{YELLOW}{cmyk}{0,0,1,0}
}

\usepackage{subfigure}
\usepackage{url}

\makeatother

\usepackage{babel}
\begin{document}

\title{The Fast Linear Accelerator Modeling Engine for FRIB Online Model
Service}

\author{Z. He}
\email{hez@frib.msu.edu}

\affiliation{Facility for Rare Isotope Beams, East Lansing, Michigan}

\author{J. Bengtsson}

\affiliation{EPIC Consulting, Ocean City, Maryland}

\author{M. Davidsaver}

\affiliation{Facility for Rare Isotope Beams, East Lansing, Michigan}

\author{K. Fukushima}

\affiliation{Facility for Rare Isotope Beams, East Lansing, Michigan}

\author{G. Shen}

\affiliation{Facility for Rare Isotope Beams, East Lansing, Michigan}

\author{M. Ikegami}

\affiliation{Facility for Rare Isotope Beams, East Lansing, Michigan}

\date{\today}
\begin{abstract}
Commissioning of a large accelerator facility like FRIB needs support
from an online beam dynamics model. Considering the new physics challenges
of FRIB such as modeling of non-axisymmetric superconducting RF cavities
and multi-charge state acceleration, there is no readily available
online beam tuning code. The design code of FRIB super-conducting
linac, IMPACT-Z, is not suitable for online tuning because of its
code design and running speed. Therefore, the Fast Linear Accelerator
Modeling Engine (FLAME), specifically designed to fulfill FRIB's online
modeling challenges, is proposed. The physics model of FLAME, especially
its novel way of modeling non-axisymmetric superconducting RF cavities
using a multipole expansion based thin-lens kick model, is discussed
in detail, and the benchmark results against FRIB design code is presented,
after which the software design strategy of FLAME and its execution
speed is presented.
\end{abstract}
\maketitle

\section{INTRODUCTION}

The Facility for Rare Isotope Beams (FRIB) is a new national user facility
for nuclear science co-funded by the Department of Energy and Michigan State University (MSU), whose driver linac is
designed to accelerate all stable ions to on-target energy larger
than 200 MeV/u and power up to 400 kW. The FRIB driver linac consists
of a front-end section where ions are generated and accelerated to 0.5
MeV/u, three linac segments with super-conducting Radio-frequency (RF) cavities where
heavy ions gain most of their kinetic energy, two folding segments
to confine the footprint and facilitate charge selection, and a beam
delivery system to transport the beam to the target~\cite{wei2013progress,wei2012frib}.
To commission and support daily operation of such a complex driver
linac with thousands of tunable components can be quite a challenging
endeavor, and an online model is essential to handle this task.

Due to the special modeling challenges, especially non-axisymmetric
superconducting RF cavities~\cite{facco2011beam,cavenago1992determining}
and multi-charge state acceleration~\cite{betz1972charge}, there
is no readily available online physics model which can handle all
the FRIB modeling challenges. And the design model of the FRIB super-conducting
linac, IMPACT-Z~\cite{qiang1999object}, is only utilized for off-line
use. In particular, it is not fast enough for online tuning. Therefore,
a physics model specially designed for online purpose while covering
all special FRIB modeling challenges with satisfactory execution speed
is desired.

The Fast Linear Accelerator Modeling Engine (FLAME) is a novel beam
envelope tracking code providing solutions to FRIB-specific challenges
of modeling non-axisymmetric superconducting RF cavities and multi-charge
state acceleration with control room quality. FLAME also implements
matrix models for other major beam lattice elements of FRIB such as
solenoid, magnetic and electrostatic quadrupole, magnetic and electrostatic
dipole, and charge stripper. Modeling of element misalignment is also
implemented. In the first section of this paper, the physics modeling
strategy is discussed. Then, the code structure design and python
interface of FLAME are discussed. Finally, calculation results of
FLAME are benchmarked against FRIB design code, and execution speed
of FLAME is also presented.

\section{Beam Dynamics Model for FLAME}

The design code for FRIB super-conducting linac, IMPACT-Z, is a Particle-In-Cell
(PIC) code. It is precise, but slow when considering online beam
tuning applications. The often used online code, OpenXAL~\cite{pelaia2015open}
for example, is based on the beam envelope formalism. However, OpenXAL
lacks some important features which are essential for correctly modeling
FRIB. FLAME is a new online code specially designed to cover all
FRIB modeling challenges. The first part of the paper describes the
physics models for FLAME.

For backwards compatitability with Trace-3D \cite{crandall1987trace,crandalltrace} the
coordinate system and units used by FLAME are 
\[
\bar{x}=[x(mm),x'(rad),y(mm),y'(rad),\phi(rad),W(MeV/u)]
\],
where $\left[x,y\right]$ are the transverse coordinates, $\left[x',y'\right]$
the derivatives, $\phi$ the longitudinal phase and $W$ the kinetic
energy. An upgrade to instead use phase-space coordinates will be
made in a later release. The beam dynamics model keeps track of the
beam's 6D center of charge and envelope of each charge state using
the transport matrix formalism.

\subsection{RF Cavity}

\begin{figure}
\includegraphics{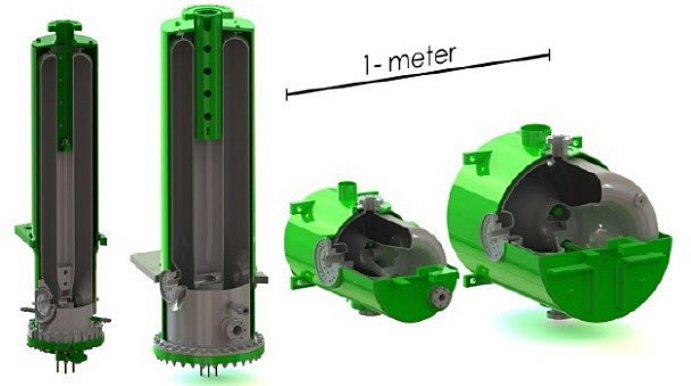} \caption{\label{Fig1}Scheme plot of 4 kinds of superconducting RF cavities
used in linac segment, namely left 1: 0.041 QWR, left 2: 0.085 QWR,
left 3: 0.29 HWR, left 4: 0.53 HWR.}
\end{figure}

FRIB uses superconducting RF cavities as the main accelerating component.
The linac segment contains 4 types of superconducting RF cavities,
namely $\beta=0.041$ quarter-wave resonator (QWR), $\beta=0.085$ QWR, $\beta=0.29$ half-wave resonator (HWR) and $\beta=0.53$ HWR, which is shown in Fig. \ref{Fig1}~\cite{facco2012superconducting}.
The physics models for superconducting RF cavities are discussed in
this section.

\subsubsection{Longitudinal Model}

For a linac, keeping track of the kinetic energy and phase evolution
of the reference synchronous particle is important. To ensure precision,
the field map integration shown in Eq. \ref{iter}
\begin{equation}
\left\{ \begin{array}{l}
W(z)=W_{s}+\int_{s}^{z}qE_{z}(z)cos\phi(z)dz\\
\phi(z)=\phi_{s}+\int_{s}^{z}\frac{2\pi}{\beta(z)\lambda}dz\\
\beta(z)=\sqrt{1-\frac{1}{\gamma(z)^{2}}}\\
\gamma(z)=\frac{W(z)}{m_{0}c^{2}}
\end{array}\right.\label{iter}
\end{equation}
is used. $z$ stands for the longitudinal position. Subscript $s$ stands for start
point. $W$ is total energy (rest energy plus kinetic energy). $q$
is charge state. $E_{z}(z)$ is longitudinal electric field map along
$z$. $\phi$ is cavity phase. $\lambda$ is speed of light over cavity
frequency. $m_{0}c^{2}$ is rest energy.

The first order longitudinal phase space transverse matrix, which
is shown in Eq. \ref{longi}, is used to track the longitudinal phase
and energy spread,
\begin{equation}
\left[\begin{array}{c}
\Delta\phi_{e}\\
\Delta W_{e}
\end{array}\right]=\left[\begin{array}{cc}
1 & 0\\
-qV_{0}Tsin\phi-qV_{0}Scos\phi & 1
\end{array}\right]\left[\begin{array}{c}
\Delta\phi_{s}\\
\Delta W_{s}
\end{array}\right]\label{longi}
\end{equation}
, where subscript $s$ stands for start point and subscript $e$ stands
for end point. $\Delta\phi$ and $\Delta W$ are RMS phase spread
and energy spread. $T$ and $S$ are transit time factors (TTFs) for
longitudinal electric field. $\phi$ is the cavtity phase at the longitudinal
focusing thin lens.

\subsubsection{Transverse Model}

As can be seen from Fig. \ref{Fig1}, these cavities are non-axisymmetric along
the beam line. It is known that non-axisymmetric super-conducting
RF cavities produce dipole and quadrupole terms in transverse directions,
and can cause beam steering and deformation~\cite{facco2011beam,cavenago1992determining}.
A multipole-expansion based scheme is implemented in FLAME.

\textbf{Field Multipole Expansion:} Multipole expansion is the traditional
method of analyzing multipole components in an electromagnetic field~\cite{he2012analytical,olave2012multipole,de2013design}.
In the 3-D cylindrical coordinate system the zeroth order of $\phi$ represents the
focusing term, the first order represents the dipole or the  steering term
and the second order represents the quadrupole term. After sampling at
a certain z, we get a 2-D vector field in the polar coordinate, then we
can transfer the 2-D vector field into two 2-D scalar fields by projection.
For each scalar field, we expand the $\rho$ direction into the Taylor
series and $\phi$ direction into the Fourier series.The coefficient is
proportional to the strength of the certain field mode. The process
can be expressed as Eq. \ref{eq2},
\begin{widetext} 
\begin{equation}
F_{\rho,nm}(\rho,\theta)=F_{max}\left[\begin{array}{c}
1\\
\rho\\
\rho^{2}\\
\rho^{3}\\
...\\
\rho^{n}
\end{array}\right]^{T}\left[\begin{array}{ccccc}
a_{00}-ib_{00} & a_{01}-ib_{01} & a_{02}-ib_{02} & ... & a_{0m}-ib_{0m}\\
a_{10}-ib_{10} & a_{11}-ib_{11} & a_{12}-ib_{12} & ... & a_{1m}-ib_{1m}\\
a_{20}-ib_{20} & a_{21}-ib_{21} & a_{22}-ib_{22} & ... & a_{2m}-ib_{2m}\\
... & ... & ... & ... & ...\\
a_{n0}-ib_{n0} & a_{n1}-ib_{n1} & a_{n2}-ib_{n2} & ... & a_{nm}-ib_{nm}
\end{array}\right]\left[\begin{array}{c}
1\\
e^{i\theta}\\
e^{i2\theta}\\
e^{i3\theta}\\
...\\
e^{im\theta}
\end{array}\right]=F_{max}\sum_{n,m=0}^{\infty}P_{n}A_{nm}\Theta_{m}\label{eq2}
\end{equation}
\end{widetext}.
Note that $\rho$ can be usually normalized by $\rho_{m}$, which
indicates the cavity aperture, to become an absolute value. $F_{\text{max}}$
is the maximum field value. As a result, $P_{n}$, $A_{nm}$ and $\Theta_{m}$
are all absolute value. Next, we choose FRIB QWR as an example to
demonstrate the above method.

E\&M fields of FRIB QWRs are simulated using Computer Simulation Technology (CST). Because the CST
numerical RF field data is in Cartesian coordinate initially, we first
transfer the E\&M field into polar coordinate by bilinear interpolation,
and then project the 2-D vector field into two 2-D scalar fields.
For the particular case of the QWR cavity, because the whole x=0 plane
is under the magnetic boundary symmetric condition, we can apply it
to simplify the Taylor-Fourier series expansion into Eq. \ref{eq3},
\begin{equation}
\begin{split}F_{\Omega,ij}=F_{max}\left[\begin{array}{c}
1\\
\rho\\
\rho^{2}\\
\rho^{3}\\
...\\
\rho^{n}
\end{array}\right]^{T}\left[\begin{array}{ccccc}
a_{00} & a_{01} & a_{02} & ... & a_{0j}\\
a_{10} & a_{11} & a_{12} & ... & a_{1j}\\
a_{20} & a_{21} & a_{22} & ... & a_{2j}\\
... & ... & ... & ... & ...\\
a_{i0} & a_{i1} & a_{i2} & ... & a_{ij}
\end{array}\right]\left[\begin{array}{c}
1\\
sin\theta\\
cos2\theta\\
sin3\theta\\
...
\end{array}\right]\\
F_{\Pi,ij}=F_{max}\left[\begin{array}{c}
1\\
\rho\\
\rho^{2}\\
\rho^{3}\\
...\\
\rho^{n}
\end{array}\right]^{T}\left[\begin{array}{ccccc}
a_{00} & a_{01} & a_{02} & ... & a_{0j}\\
a_{10} & a_{11} & a_{12} & ... & a_{1j}\\
a_{20} & a_{21} & a_{22} & ... & a_{2j}\\
... & ... & ... & ... & ...\\
a_{i0} & a_{i1} & a_{i2} & ... & a_{ij}
\end{array}\right]\left[\begin{array}{c}
0\\
cos\theta\\
sin2\theta\\
cos3\theta\\
...
\end{array}\right]
\end{split}
\label{eq3}
\end{equation},
where $\Omega$ means $E_{\rho}$ (electric field along radial direction)
or $H_{\theta}$ (magnetic field along azimuthal direction), $\Pi$
means $E_{\theta}$ (electric field along azimuthal direction) or
$H_{\rho}$ (magnetic field along radial direction). $a_{ij}$ is
the multipole coefficient.

To calculate the coefficient of each multipole term, we can make use
of the orthogonality of the trigonometric function and integrate
with $\sin(k\theta)$ or $\cos(k\theta)$. $k$ is the order number. Then we expand the result
into the Taylor series of $\rho$. The coefficient of each
order corresponds to the relative strength of the multipole term.

Theoretically, we should expand the field to an infinite order, while in practice,
we have to do truncation at a certain order. In principle, we can
truncate the expansion to any order as long as all important multipoles
have been included. Here, we truncate both the Taylor and the Fourier series
to include all linear terms, namely, focusing, dipole and quadrupole
terms.

By sampling along the longitudinal direction and calculating multipole
coefficients, we can have curves that indicate the multipole field strength
versus the longitudinal position. The following Fig. \ref{Fig3} shows
the multipole strength curve for the 0.085 QWR electric-magnetic field.

\begin{figure}
\begin{minipage}[t]{0.5\linewidth}%
 \centering \subfigure{ \label{Fig3a} 
\includegraphics[width=1.7in,height=1.5in]{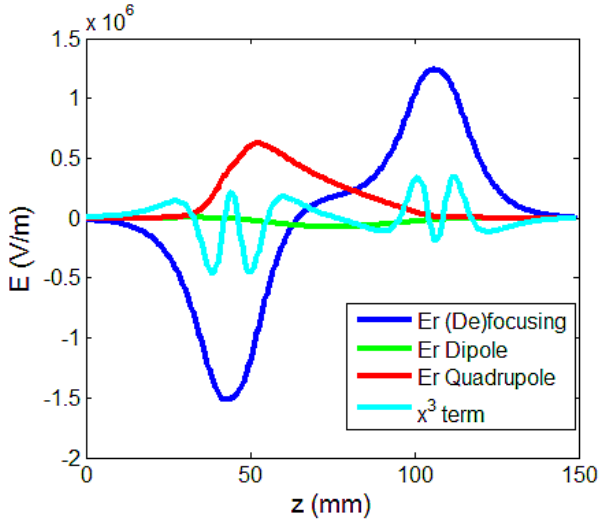} } %
\end{minipage}%
\begin{minipage}[t]{0.5\linewidth}%
 \subfigure{ \label{Fig3b} 
\includegraphics[width=1.7in,height=1.5in]{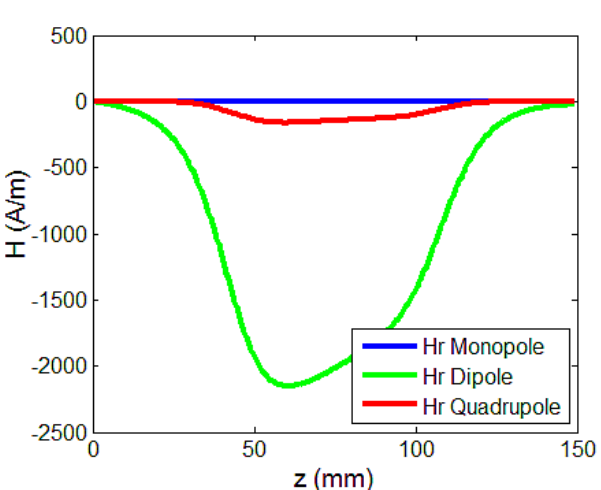} } %
\end{minipage}\caption{Results of multipole strength curves for radial electric and magnetic
fields. (a) The radial electric field multipole strength, defined as $E_{max}$
times multipole term coefficient, v.s. longitudinal coordinate z. Blue
curve is (de)focusing term, green curve is dipole term, red curve is quadrupole
term, cyan curve is cubic term. (b) The radial magnetic field multipole
strength, defined as $H_{max}$ times multipole term coefficient,
v.s. longitudinal coordinate $z$. Blue curve is (de)focusing term, green
curve is dipole term, and red curve is quadrupole term. }

\label{Fig3} 
\end{figure}

\textbf{The Multipole Beam Kick Model:} After getting out the coefficient
matrix from Taylor-Fourier expansion, the multipole beam kick model
is used to calculate the beam kick caused by multipole terms. We start
from the Lorentz force equation and assume a small acceleration to 
calculate the change in $y'$ in the vertical direction, shown in
Eq. \ref{eq4},
\begin{equation}
\small\begin{split} & \Delta y'=\frac{qe\mu_{0}}{\gamma m_{0}}\int_{t_{1}}^{t_{2}}H_{x}(x,y,z,t)dt+\frac{qe}{\gamma m_{0}\beta c}\int_{t_{1}}^{t_{2}}E_{y}(x,y,z,t)dt\\
 & =\Delta y'_{H,y}+\Delta y'_{E,y}
\end{split}
\label{eq4}
\end{equation},
where $q$ is the number of charge, $e$ is the elementary charge amount,
$m_{0}$ is the particle static mass, $\mu_{0}$ is the permeability,
$c$ is the speed of light. The beam kick can be divided into the magnetic
kick and the electric kick. Both kicks can be expressed with a function
of $x,y,z$ and $t$. Take the electric kick as an example. If we assume the field
is a sinusoidal function about time, then we use $z$ instead of t as an independent
variable, and then we get Eq. \ref{eq5},
\begin{equation}
\Delta y'_{E,y}=\frac{qe}{\gamma\beta^{2}m_{0}c^{2}}\int_{z_{1}}^{z_{2}}E_{y}(x,y,z)cos(kz+\phi_{0})dz\label{eq5}
\end{equation},
where $\phi_{0}$ is the cavity phase when the particle is at the entrance of the
cavity. After transformation and simplification,
we can write $E_{y}$ into the following form shown in Eq. \ref{eq6},
\begin{equation}
E_{y}(x,y,z)=\sum_{i,j=0}^{n}E_{max}(z)a_{ij}(z)t_{ij}(x,y)\label{eq6}
\end{equation}.
$E_{max}(z)$ is the maximum electric field at the 2-D plane with the
longitudinal position $z$. $a_{ij}(z)$ is the multipole coefficient
after coordinate transformation. $t_{ij}(x,y)$ is the coordinate transferring
factor.

We can substitute it into Eq. \ref{eq4} and make use of the concept of transit
time factors, and get the expression of the vertical beam kick
by electric multipole components,
\begin{equation}
\Delta y'_{E,y}=\frac{qe}{\gamma\beta^{2}m_{0}c^{2}}\sum_{i,j=0}^{n}t_{ij}V_{ij}(T_{ij}cos\phi-S_{ij}sin\phi)\label{eq7}
\end{equation}
,where $V_{ij}$ is the voltage of the multipole term $i,j$ defined by integration of multipole term strength $E_{max}A_{ij}(z)$ along longitudinal direction with a unit
of Volt. $T_{ij}$ and $S_{ij}$ are the transit
time factors of multipole term $i,j$. $\phi$ is the cavity phase
at the reference point.

Similarly, we can also get the expression of vertical beam kick by
magnetic multipole components,
\begin{equation}
\Delta y'_{H,y}=\frac{qe\mu_{0}}{\gamma\beta m_{0}c^{2}}\sum_{i,j=0}^{n}t_{ij}U_{ij}(T_{ij}cos\phi-S_{ij}sin\phi)\label{eq8}
\end{equation}.
The definition of $U_{ij}$ is similar to $V_{ij}$ except the electric field is changed 
to the magnetic field, so the unit becomes Ampere.

The entire procedure can be repeated for the calculation of horizontal case.

\textbf{The Focusing Term:} The term with $i=1,j=0$ stands for the beam focusing
term. For this case, $E_{max}(z)=E_{max,\rho}(z)$
or the maximum electric field in a 2D plane equals the maximum electric field
of the radial projection while the azimuthal projection of the electric field
for a focusing term is zero. The coordinate transferring factor $t_{ij}=y/\rho_{m}$.
For magnetic field case, the (de)focusing term $H_{max}(z)=H_{max,\theta}(z).t_{ij}=-y/\rho_{m}$.
The calculation of the multipole expansion sees very little magnetic (de)focusing
term in QWRs and HWRs, so focusing coming from the magnetic (de)focusing term
usually can be omitted. Using the model, we can evaluate the focusing or
defocusing effect of the cavity.

\textbf{The Beam Steering Term:} The term with $i=0,j=1$ stands for the beam steering
term, or the dipole term. For the electric field, $E_{max}(z)=E_{max,\rho}(z)=E_{max,\theta}(z)$,
and the coordinate transferring factor $t_{ij}=1$. For the magnetic field,
$H_{max}(z)=H_{max,\rho}(z)=H_{max,\theta}(z)$, and $t_{ij}=1$.
The calculation of the multipole expansion sees both significant electric
and magnetic dipole terms, so we expect that the beam dipole kick can
come from both sources.

\begin{figure}
\centering \subfigure{ \label{Fig4a} 
\includegraphics[width=3.4in,height=1.8in]{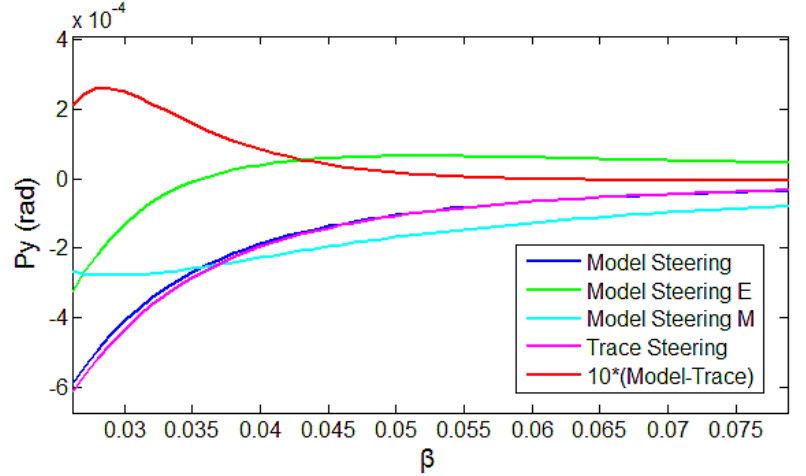} } \subfigure{
\label{Fig4b} 
\includegraphics[width=3.4in,height=1.8in]{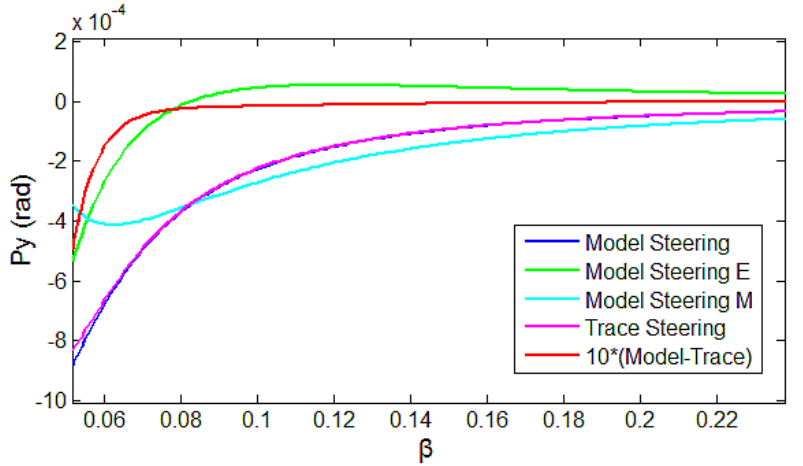} } \caption{Prediction of beam steering by model and tracking vs. $\beta$, synchronous
phase $\phi_{s}=-\pi/6$. Blue curve shows the total beam steering
calculated by the model, the green and cyan curve each stands for the electric
field steering and the magnetic field steering by the model, the magenta curve
shows result calculated from tracking, and the red curve shows 10 times
the error between the model and tracking; (a) 0.041 QWR; (b) 0.085 QWR. }

\label{Fig4} 
\end{figure}

Using the multipole beam kick model described above, we can calculate
the dipole kick and predict beam steering of FRIB QWRs. The results
can be benchmarked with particle tracking results, shown in Fig. \ref{Fig4}.
Both FRIB 0.041 QWR and 0.085 QWR are used. The synchronous phase is fixed
at $-\pi/6$, and the accelerating voltage is the design voltage. The green
curve indicates the beam steering effect coming from the electric dipole
while the cyan curve indicates the beam steering coming from the magnetic
dipole. The blue curve is the combination of these two effects and
is benchmarked well against the result from 3-D field particle
tracking which is plotted in magenta. The difference between the model
and 3-D field tracking is plotted in red, note that in the figure
the error value is magnified by 10.

We can see that the precision of the model is high, especially in the high
energy region. In the low energy region, particle acceleration in a cavity
cannot be neglected so that the error goes up. When the accelerating
field of a QWR cavity is fixed, the steering effect would be damping
with $\beta$. It is because the higher the energy is, the harder
it can be bent by both the electric and magnetic field. Also, damping
of the electric field bending effect is faster than that of the magnetic
field. So in the high energy region, the dipole steering effect becomes less
important and magnetic dipole would be a dominating term. In the low energy
region, the dipole steering could become a quite serious problem and some
measures should be taken to correct it.

\textbf{The Quadrupole Term:} The term with $i=1,j=2$ stands for the quadrupole
term. For the vertical electric field, $E_{max}(z)=E_{max,\rho}(z)=E_{max,\theta}(z)$,
the coordinate transferring factor is $t_{ij}=y/\rho_{max}$. For the magnetic
field, $H_{max}(z)=H_{max,\rho}(z)=H_{max,\theta}(z)$ , $t_{ij}=y/\rho_{max}$.
We can also use the model described above to calculate the quadrupole
kick, and then benchmark the calculated quadrupole strength against particle
tracking.

\begin{figure}
\centering \subfigure{ \label{Fig5a} 
\includegraphics[width=3.4in,height=1.8in]{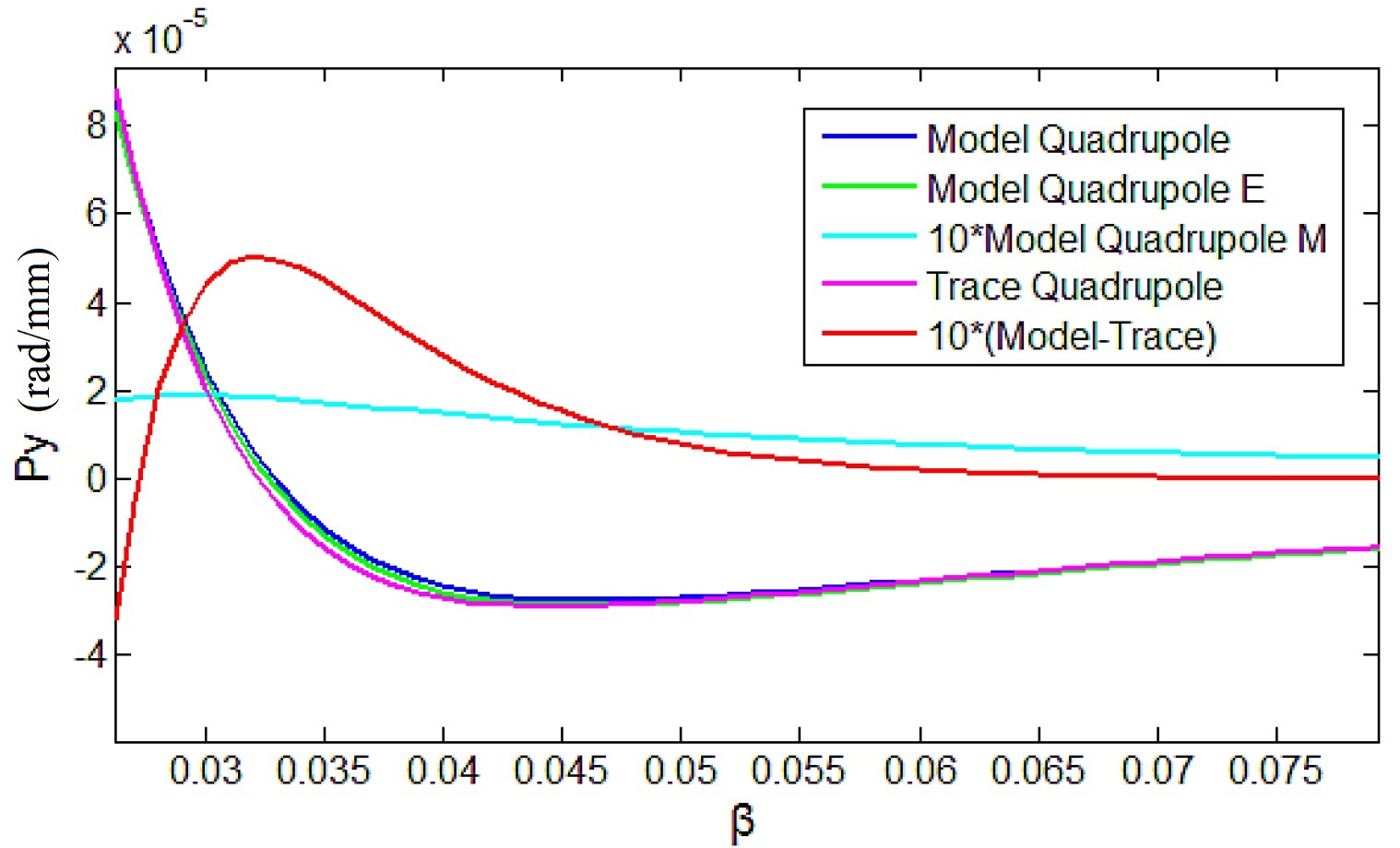} } \subfigure{
\label{Fig5b} 
\includegraphics[width=3.4in,height=1.8in]{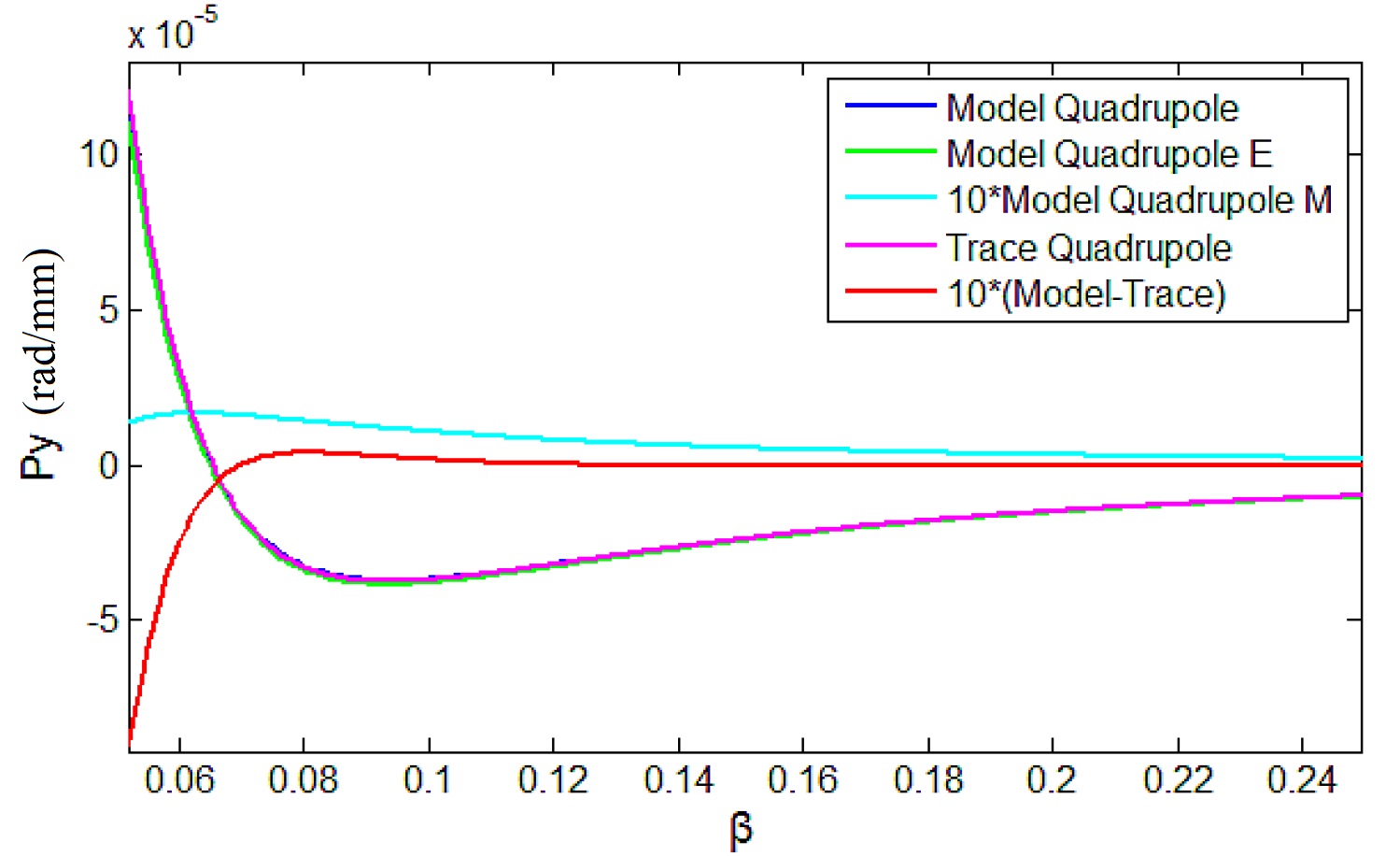} } \caption{Prediction of quadrupole strength by the model and tracking vs. $\beta$,
the synchronous phase $\phi_{s}=-\pi/6$. The blue curve shows the total quadrupole
strength calculated by the model, the green and cyan curve each stands for
the electric field quadrupole and 10 times magnified magnetic field quadrupole by
the model, the magenta curve shows result calculated from tracking, and the red
curve shows 10 times the error between the model and tracking. (a) 0.041
QWR; (b) 0.085 QWR.}

\label{Fig5} 
\end{figure}

Results can be seen in Fig. \ref{Fig5}. Both the FRIB 0.041 QWR and 0.085
QWR are used. The synchronous phase is fixed at $-\pi/6$, the accelerating
voltage is the design voltage. The green curve stands for the quadrupole
kick coming from the electric field and the cyan curve stands for the quadrupole
kick coming from the magnetic field (10 times magnified). The blue curve
is the total quadrupole kick of the electric and magnetic field
contribution added together. We can also see good agreement when comparing
the quadrupole kick predicted by the model (blue curve) against the prediction
from particle tracking (magenta curve). The error (10 times magnified)
is shown in the red curve. According to the model, we can draw the conclusion
that the quadrupole kick mainly comes from the electric field and there is
nearly no contribution from the magnetic field. The quadrupole effect would
also be damping with $\beta$ growing and the effect would be more
significant when at low $\beta$.

\begin{figure}
\begin{minipage}[t]{0.5\linewidth}%
 \centering \subfigure{ \label{Fig6a} 
\includegraphics[width=1.7in,height=1.5in]{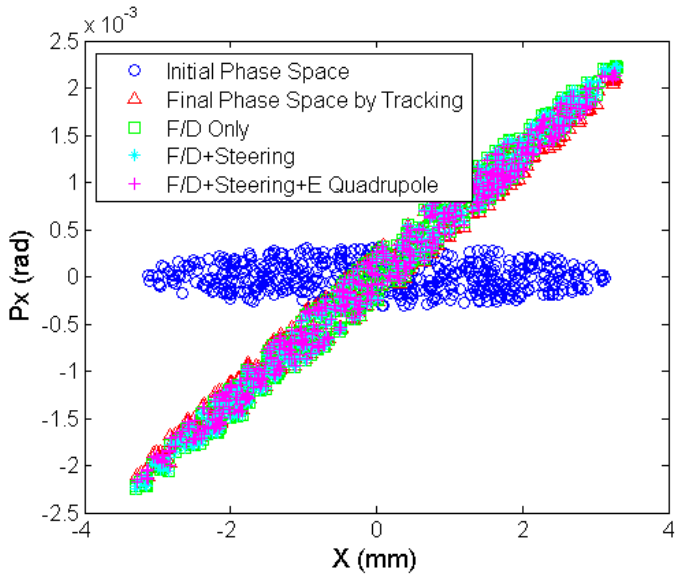} } %
\end{minipage}%
\begin{minipage}[t]{0.5\linewidth}%
 \subfigure{ \label{Fig6b} 
\includegraphics[width=1.7in,height=1.5in]{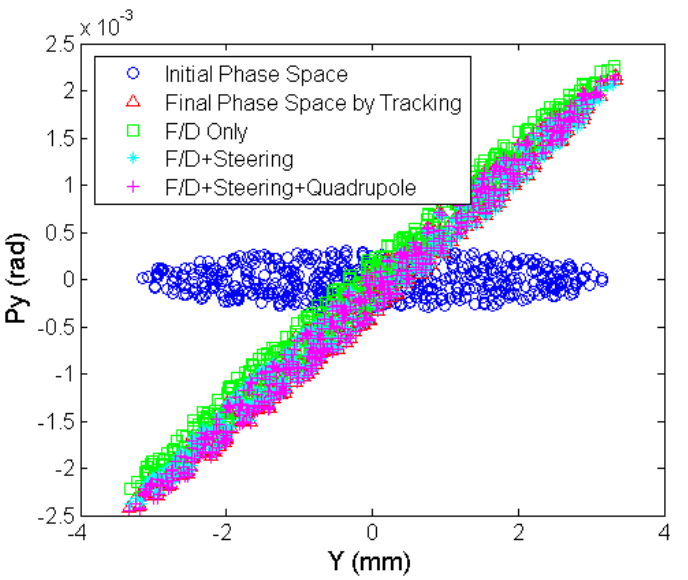} } %
\end{minipage}\ %
\begin{minipage}[t]{0.5\linewidth}%
 \centering \subfigure{ \label{Fig6c} 
\includegraphics[width=1.7in,height=1.5in]{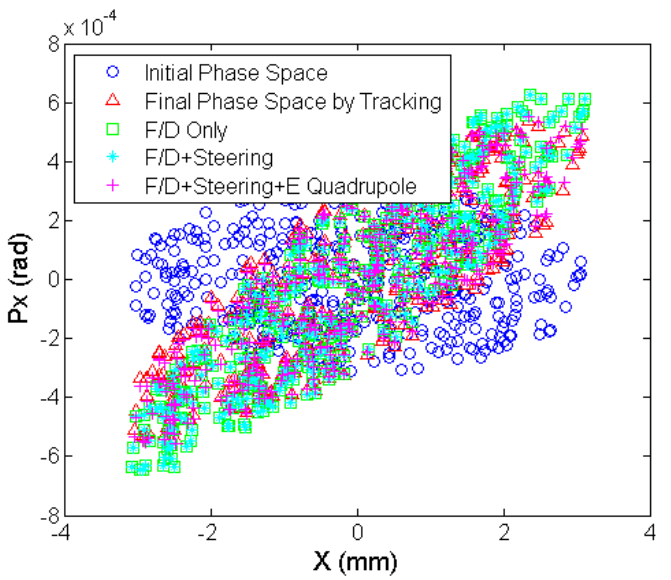} } %
\end{minipage}%
\begin{minipage}[t]{0.5\linewidth}%
 \subfigure{ \label{Fig6d} 
\includegraphics[width=1.7in,height=1.5in]{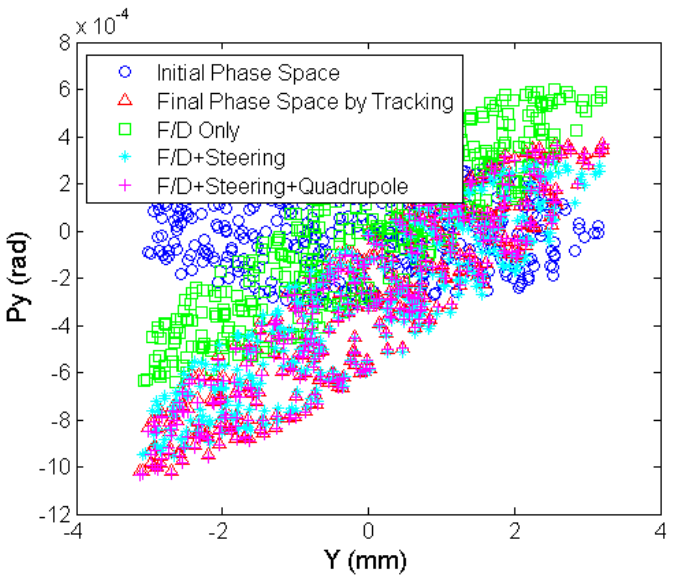} } %
\end{minipage}\caption{The horizontal phase space and the vertical phase space of a KV distribution beam using tracking and models including different field terms; (a) horizontal phase
space, 0.041 QWR; (b) vertical phase space, 0.041 QWR; (c) horizontal phase space,
0.085 QWR; (d) vertical phase space, 0.085 QWR.}

\label{Fig6} 
\end{figure}

\textbf{The Transverse Thin Lens Model}: The transverse drift-kick-drift
thin lens model is then developed. The thin lens model includes a
series of multipole kicks located at the center of the components,
and are separated by a series of drift spaces. For different components,
the transfer maps for the transverse phase space are shown as below:

For electric (de)focusing,
\begin{equation}
\begin{split} & \left\{ \begin{array}{l}
x_{2}=x_{1}\\
x_{2}'=x_{1}'+K_{0}x_{1}
\end{array}\right.\left\{ \begin{array}{l}
y_{2}=y_{1}\\
y_{2}'=y_{1}'+K_{0}y_{1}
\end{array}\right.\\
 & K_{0}=\frac{qe}{\beta^{2}\gamma m_{0}c^{2}\rho_{m}}V_{E,0}(T_{E,0}cos\phi-S_{E,0}sin\phi)
\end{split}
\label{eq9}
\end{equation}.
For the electric dipole,
\begin{equation}
\begin{split} & \left\{ \begin{array}{l}
x_{2}=x_{1}\\
x_{2}'=x_{1}'
\end{array}\right.\left\{ \begin{array}{l}
y_{2}=y_{1}\\
y_{2}'=y_{1}'+\Delta y'
\end{array}\right.\\
 & \Delta y'=\frac{qe}{\beta^{2}\gamma m_{0}c^{2}}V_{E,1}(T_{E,1}cos\phi-S_{E,1}sin\phi)
\end{split}
\label{eq10}
\end{equation}.
For the electric quadrupole,
\begin{equation}
\begin{split} & \left\{ \begin{array}{l}
x_{2}=x_{1}\\
x_{2}'=x_{1}'+K_{q}x_{1}
\end{array}\right.\left\{ \begin{array}{l}
y_{2}=y_{1}\\
y_{2}'=y_{1}'-K_{q}y_{1}
\end{array}\right.\\
 & K_{q}=\frac{qe}{\beta^{2}\gamma m_{0}c^{2}\rho_{m}}V_{E,2}(T_{E,2}cos\phi-S_{E,2}sin\phi)
\end{split}
\label{eq11}
\end{equation}.
For magnetic (de)focusing,
\begin{equation}
\begin{split} & \left\{ \begin{array}{l}
x_{2}=x_{1}\\
x_{2}'=x_{1}'+K_{0}x_{1}
\end{array}\right.\left\{ \begin{array}{l}
y_{2}=y_{1}\\
y_{2}'=y_{1}'+K_{0}y_{1}
\end{array}\right.\\
 & K_{0}=\frac{qec\mu_{0}}{\beta\gamma m_{0}c^{2}\rho_{m}}U_{H,0}(T_{H,0}cos\phi-S_{H,0}sin\phi)
\end{split}
\label{eq12}
\end{equation}.
For the magnetic dipole,
\begin{equation}
\begin{split} & \left\{ \begin{array}{l}
x_{2}=x_{1}\\
x_{2}'=x_{1}'
\end{array}\right.\left\{ \begin{array}{l}
y_{2}=y_{1}\\
y_{2}'=y_{1}'+\Delta y'
\end{array}\right.\\
 & \Delta y'=\frac{qec\mu_{0}}{\beta\gamma m_{0}c^{2}}U_{H,1}(T_{H,1}cos\phi-S_{H,1}sin\phi)
\end{split}
\label{eq13}
\end{equation}.
For the magnetic quadrupole,
\begin{equation}
\begin{split} & \left\{ \begin{array}{l}
x_{2}=x_{1}\\
x_{2}'=x_{1}'+K_{q}x_{1}
\end{array}\right.\left\{ \begin{array}{l}
y_{2}=y_{1}\\
y_{2}'=y_{1}'-K_{q}y_{1}
\end{array}\right.\\
 & K_{q}=\frac{qec\mu_{0}}{\beta\gamma m_{0}c^{2}\rho_{m}}U_{H,2}(T_{H,2}cos\phi-S_{H,2}sin\phi)
\end{split}
\label{eq14}
\end{equation}.
Subscripts $E,0$, $E,1$ and $E,2$ mean electric (de)focusing, dipole
and quadrupole terms. Subscripts $H,0$, $H,1$ and $H,2$ mean magnetic
(de)focusing, dipole and quadrupole terms. Then we use the thin lens model
for tracking, and the result is benchmarked against 3-D field particle
tracking. Both the 0.041 QWR and the 0.085 QWR are tested. The design beta
and the design accelerating voltage are used, and the synchronous
phase is fixed at $-\pi/6$. The result is shown in Fig. \ref{Fig6}.
Blue circles represent the initial phase space and red triangles represent
the final phase space calculated by 3-D field particle tracking. Green
squares show the thin lens model with focusing components only, which
tilt both the x and y phase space. Blue stars represent the phase space
after adding the steering term. There is no steering effect in the x direction,
while in the y direction the steering is about -0.176 mrad for the 0.041 QWR
and -0.245 mrad for the 0.085 QWR at design $\beta$. The RMS errors of
blue stars for $x,x',y,y'$ are 0.42\%, 5.28\%, 0.75\%, 2.20\% for
the 0.041 QWR, and 1.45\%, 23.91\%, 1.39\%, 9.17\% for the 0.085 QWR respectively.
In the figure, red crosses represent the phase space after adding
quadrupole terms. Then the x and y phase spaces become tilted and get
closer to the 3-D field particle tracking. The RMS errors of red crosses
for $x,x',y,y'$ become 0.16\%, 1.83\%, 0.23\%, 1.59\% for the 0.041
QWR and 0.07\%, 4.59\%, 0.27\%, 2.41\% for the 0.085 QWR respectively.

\subsection{Solenoid}

Superconducting solenoids are used as main focusing elements within
cryomodules in the FRIB driver linac. A solenoid is modeled by the transport matrix described in Eq. \ref{sole} ~\cite{crandalltrace},
\begin{equation}
M_{\text{sol}}=\left[\begin{array}{cccc}
C^{2} & \frac{SC}{g} & SC & \frac{S^{2}}{g}\\
-gSC & C^{2} & -gS^{2} & SC\\
-SC & -\frac{S^{2}}{g} & C^{2} & \frac{SC}{g}\\
gS^{2} & -SC & -gSC & C^{2}
\end{array}\right]\label{sole}
\end{equation},
where $C=\cos\left(\theta\right)$, $S=\sin\left(\theta\right)$,
$g=\frac{B}{2\left(B\rho\right)}$ is the focusing strength , $B$
the magnetic field, $\left(B\rho\right)=\frac{p}{q}$ the particle's
magnetic rigidity, $q$ the charge state, $p$ the momentum, $\theta$
is $gL$. $L$ is the solenoid's length. For a realistic solenoid, the effective magnetic field strength and 
length is,
\begin{equation}
\hat{B_{z}}=\frac{\int_{-\infty}^{\infty}B_{z}^{2}(z)}{\int_{-\infty}^{\infty}B_{z}(z)},\hat{L}=\frac{[\int_{-\infty}^{\infty}B_{z}(z)]^{2}}{\int_{-\infty}^{\infty}B_{z}^{2}(z)}\label{SoleEff}
\end{equation}.

\subsection{Quadrupole}

Room temperature quadrupoles are also used as focusing element outside
cryomodules. The 2\texttimes 2 quadrupole transport matrices (Eq.
\ref{quad}) ~\cite{crandalltrace} for a quadrupole with magnetic
field gradient $B'$ in the focusing and defocusing planes are,
\begin{equation}
\left\{ \begin{array}{l}
F=\left[\begin{array}{cc}
\cos\left(\sqrt{k}L\right) & \frac{1}{\sqrt{k}}\sin\left(\sqrt{k}L\right)\\
-k\sin\left(\sqrt{k}L\right) & \cos\left(\sqrt{k}L\right)
\end{array}\right]\\
D=\left[\begin{array}{cc}
\cosh\left(\sqrt{k}L\right) & \frac{1}{\sqrt{k}}\sinh\left(\sqrt{k}L\right)\\
k\sinh\left(\sqrt{k}L\right) & \sinh\left(\sqrt{k}L\right)
\end{array}\right]
\end{array}\right.\label{quad}
\end{equation}
\begin{flushleft},
where $k=\left|\frac{B'}{\left(B\rho\right)}\right|$ is the focusing
strength.
\par\end{flushleft}

\subsection{Dipole}

The dipole transport matrix is Eq. \ref{bend} ~\cite{brown1973transport} for
$\left[x,x',y,y',l,\delta\right]$,
\begin{equation}
M_{\text{bend}}=\left[\begin{array}{cccccc}
C & \frac{S}{h_{0}} & 0 & 0 & 0 & \frac{1-C}{h_{0}}\\
-h_{0}S & C & 0 & 0 & 0 & S\\
0 & 0 & 1 & L & 0 & 0\\
0 & 0 & 0 & 1 & 0 & 0\\
S & \frac{1-C}{h_{0}} & 0 & 0 & 1 & \alpha_{\mathrm{c}}L\\
0 & 0 & 0 & 0 & 0 & 1
\end{array}\right]\label{bend}
\end{equation},
where $C=\cos\left(\alpha\right)$, $S=\sin\left(\alpha\right)$,
$h_{0}=\frac{1}{\rho_{0}}$, $\alpha$ the bend angle, $\rho_{0}$
the bend radius, and $\alpha_{\mathrm{c}}$ the momentum compaction,
\begin{equation}
\alpha_{\mathrm{c}}=1-\frac{S}{Lh_{0}}
\end{equation}.
 For the coordinates used by FLAME $\left[x,x',y,y',\phi,W\right]$
, the dipole transport matrix is
\begin{equation}
M_{\text{bend}}=\left[\begin{array}{cccccc}
C & \frac{S}{h_{0}} & 0 & 0 & 0 & \frac{1-C}{h}\frac{1}{\left(pv\right)_{0}}\\
-h_{0}S & C & 0 & 0 & 0 & S\frac{1}{\left(pv\right)_{0}}\\
0 & 0 & 1 & L & 0 & 0\\
0 & 0 & 0 & 1 & 0 & 0\\
Sk_0 & \frac{1-C}{h_{0}}k_0 & 0 & 0 & 1 & k_{0}\eta_{\mathrm{c}}\frac{L}{\left(pv\right)_{0}}\\
0 & 0 & 0 & 0 & 0 & 1
\end{array}\right]
\end{equation},
where $\left(pv\right)_{0}$ is the momentum and velocity for the
reference particle, $k_{0}=\frac{2\pi}{\beta_{0}\lambda_{\mathrm{RF}}}$, 
$\beta_{0}$ and $ \gamma_{0}$ are the relativistic factors, and $\eta_{\mathrm{c}}$
is the phase-slip factor,
\begin{equation}
\eta_{\mathrm{c}}=1-\frac{S}{Lh_{0}}-\frac{1}{\gamma_{0}^{2}}
\end{equation}.
The transport matrix for edge focusing Eq. \ref{face} is
\begin{equation}
M_{\text{face}}=\left[\begin{array}{cccccc}
1 & 0 & 0 & 0 & 0 & 0\\
h\tan\left(\beta\right) & 1 & 0 & 0 & 0 & 0\\
0 & 0 & 1 & 0 & 0 & 0\\
0 & 0 & -h\tan(\beta-\psi) & 1 & 0 & 0\\
0 & 0 & 0 & 0 & 1 & 0\\
0 & 0 & 0 & 0 & 0 & 1
\end{array}\right]\label{face}
\end{equation},
where $\beta$ is the pole face angle, $\psi$ is the correction term
resulting from the spatial extent of fringing fields defined as $\psi=K_{1}(\frac{g}{\rho_{0}})(\frac{1+\sin{}^{2}\left(\beta\right)}{\cos\left(\beta\right)})[1-K_{1}K_{2}(\frac{g}{\rho_{0}})\tan\left(\beta\right)]$.
$K_{1}$ and $K_{2}$ are the first and second fringe field integral.
The full magnet gap width is $g$.

\subsection{Electrostatic Element}

The front-end section of FRIB uses electrostatic quadrupoles and
bends. Beam dynamic models for electrostatic quadrupoles and bends
are also included in FLAME. The transport matrix for electrostatic
quadrupoles are similar to magnetic quadrupoles. One just need to
change the definition of the focusing strength into Eq. \ref{Efc}~\cite{takeda2004parmila},
\begin{equation}
k=\frac{2V_{0}}{\left(E\rho\right)R^{2}}\label{Efc}
\end{equation},
where $\left(E\rho\right)=$$\frac{pv}{q}$ is the particle's electric
rigidity, $q$ is the charge state, $v$ is the velocity, $V_{0}$ is the voltage
of the electrode, and $R$ is the radius.

The transport matrix for a hard edge electrostatic bend is given by~\cite{wollnik2012optics} and for the coordinates used by FLAME,
\begin{equation}
M_{\text{eb}}=\left[\begin{array}{cccccc}
C_{x} & \frac{S_{x}}{\sqrt{k_{x}}} & 0 & 0 & 0 & D_{x}N_{W}\frac{1}{W_{0}}\\
-\sqrt{k_{x}}S_{x} & C_{x} & 0 & 0 & 0 & D'_{x}N_{W}\frac{1}{W_{0}}\\
0 & 0 & C_{y} & \frac{S_{y}}{\sqrt{k_{y}}} & 0 & 0\\
0 & 0 & \sqrt{k_{y}}S_{y} & C_{y} & 0 & 0\\
k_0D'_{x}N_{t} & k_0D_xN_t & 0 & 0 & 1 & k_{0}\eta_{\mathrm{c}}\frac{L}{W_{0}}\\
0 & 0 & 0 & 0 & 0 & 1
\end{array}\right]\label{EB}
\end{equation},
where $C_{x,y}=\cos\left(\sqrt{k_{x,y}}L\right)$, $S_{x,y}=\sin\left(\sqrt{k_{x,y}}L\right)$,
$D_{x}=\frac{1-\cos\left(\sqrt{k_{x}}L\right)}{\rho_{0}k_{x}}$, $D'_{x}=\frac{\sin\left(\sqrt{k_{x}}L\right)}{\rho_{0}\sqrt{k_{x}}}$,
$k_{0}=\frac{2\pi}{\beta_{0}\lambda_{\mathrm{RF}}}$, $N_{W}=\frac{\gamma_{0}^{2}+1}{\gamma_{0}\left(\gamma_{0}+1\right)}$,
and $N_{t}=\frac{\gamma_0^2+1}{\gamma_0^2}$.
$\beta_0$, $\gamma_{0}$ are the Lorentz factors. $W_0$ is the particle energy. $\lambda_{\mathrm{RF}}$ is the RF wave length. $k_{x}=\frac{1-n_{1}+\gamma_{0}^{2}}{\rho_{0}^{2}}$,
$k_{y}=\frac{n_{1}}{\rho_{0}^2}$. For a spherical or cylindrical electrostatic
bend, $n_{1}=1$ or $n_{1}=0$ respectively. The bend radius is $\rho_{0}$. $K_{0}$ is the kinetic
energy of the reference particle, and $\eta_{\mathrm{c}}$ the phase-slip factor
\begin{equation}
\eta_{\mathrm{c}}=\frac{\sqrt{k_{x}}L-S_{x}}{L\rho_{0}^{2}k_{x}^{\nicefrac{3}{2}}}N_{W}N_{t}-\frac{1}{\gamma_{0}\left(\gamma_{0}+1\right)}
\end{equation}.

The transport matrices for edge focusing are
\begin{equation}
\begin{array}{l}
M_{\text{entry}}=\left[\begin{array}{cccccc}
1 & 0 & 0 & 0 & 0 & 0\\
-\frac{1}{f'_x} & \frac{1}{\sqrt{1+\kappa}} & 0 & 0 & 0 & 0\\
0 & 0 & 1 & 0 & 0 & 0\\
0 & 0 & -\frac{1}{f'_y} & \frac{1}{\sqrt{1+\kappa}} & 0 & 0\\
0 & 0 & 0 & 0 & 1 & 0\\
0 & 0 & 0 & 0 & 0 & \frac{1}{1+\kappa}
\end{array}\right]\\
M_{\text{exit}}=\left[\begin{array}{cccccc}
1 & 0 & 0 & 0 & 0 & 0\\
-\frac{1}{f''_x} & \sqrt{1+\kappa} & 0 & 0 & 0 & 0\\
0 & 0 & 1 & 0 & 0 & 0\\
0 & 0 & -\frac{1}{f''_y} & \sqrt{1+\kappa} & 0 & 0\\
0 & 0 & 0 & 0 & 1 & 0\\
0 & 0 & 0 & 0 & 0 & 1+\kappa
\end{array}\right]
\end{array}
\label{EB}
\end{equation},
where $f'_x,f'_y,f''_x,f''_y$ parameterize the edge focal length; for the details see the~\cite{wollnik2012optics}. The kinetic energy increase factor $\kappa$ is
\begin{equation}
\kappa=\frac{W-W_0}{W_0}
\end{equation},
which characterizes the effect of particle kinetic energy change due to middle point potential deviation from ground. 

The full transport matrix is
\begin{equation}
M_{\text{ebf}}=M_{\text{exit}}M_{\text{eb}}M_{\text{entry}}
\end{equation}.

\subsection{Charge Stripper}

FRIB is using a charge stripper in folding segment 1 to transfer heavy
ions into higher charge states in order to increase the acceleration
efficiency~\cite{marti2010development}. The interaction between the beam and the charge stripper
has been simulated with SRIM~\cite{ziegler2010srim} and
then a stripper model has been built by parameterizing the probabilistic
distribution within IMPACT-Z~\cite{gorelov2006ion}. A stripper model
is also needed inside FLAME in order to complete beginning to end simulation
for FRIB.

There are three main effects which need to be modeled for a stripper:
1) change in charge states, 2) change in the beam energy and 3) blow up
in the phase space. The distribution of charge states after the stripper can be
calculated using Baron's formula shown in Eq. \ref{Baron}~\cite{baron1988beam,baron1993charge},
\begin{equation}
\left\{ \begin{array}{l}
\frac{\bar{Q}}{Z}=1-exp(-83.275\frac{\beta}{Z^{0}.447})\\
Q_{\text{ave}}=\bar{Q}(1-exp(-12.905+0.2124Z-0.00122Z^{2}))\\
d=\sqrt{\bar{Q}(0.07535+0.19Y-0.2654Y^{2})},Y=\bar{Q}/Z
\end{array}\right.\label{Baron}
\end{equation}.
Here, Gaussian distribution of output charge states is assumed. $Q_{ave}$
is the average output charge state and d is the standard deviation. $\beta$
is the incident beam Lorentz factor, and $Z$ is the number
of protons of the incident beam. 

The energy change effect and the phase space blow up effect can be modeled
using SRIM. SRIM is a Monte-Carlo code which simulates
features of the transport of ions in matter. To build a stripper model
for FRIB, we simulated a scenario where an incident 16.623 MeV/u
uranium beam with zero emittance is assumed and hit onto a 3 \textmu m
carbon stripper. Then, the energy loss and scattering angle are calculated
with SRIM. The result can be fitted by Eq.\ref{SRIMfit},
\begin{equation}
f(\theta,E)=\frac{\lambda^{\alpha}}{\Gamma(\alpha)}\theta^{\alpha-1}e^{-\lambda\theta}e^{-\frac{1}{2}[\frac{E-E_{0}}{E_{1}}]^{2}}\label{SRIMfit}
\end{equation}.
We assume Gamma distribution for the scattering angle and Gaussian distribution
for kinetic energy scattering. $\alpha$ and $\lambda$ are parameters
related to the shape of the angular distribution, $E_{0}$ is the average beam energy
loss after the stripper, and $E_{1}$ is the energy loss distribution width. To
simplify the model, we assume $E_{0}$ has no angular dependence.
The fitted result can be seen in Fig. \ref{Angfit} below. More cases
with different carbon stripper thickness and incident beam energy
are simulated, so that we can get fitting parameters over a certain
range of stripper thickness and incident beam energy through interpolation.

\begin{figure}
\centering \includegraphics[width=2.5in,height=2in]{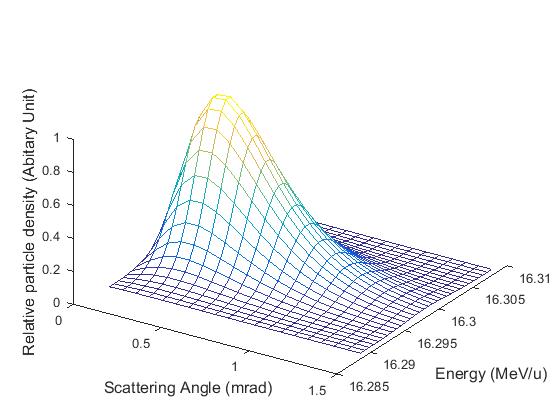}
\caption{Energy and scattering angle distribution after an incident 16.623
MeV/u zero emittance uranium beam travels through a 3 \textmu m carbon stripping
foil. The three axes are the scattering angle, the output energy, and the relative
particle density respectively.}
\label{Angfit} 
\end{figure}

After that, we can model the phase space blow up effect. We assume that the
beam does not change in the real space. The overall transverse angular spread can
be treated as an extra transverse angular spread, introduced by stripper, 
adding up to the original angular spread, which can be approximated
by Eq. \ref{AngSp},
\begin{equation}
\sigma_{h}=\sqrt{\sigma_{f}^{2}+\sigma_{\Gamma}^{2}}\label{AngSp}
\end{equation},
where $\sigma_{f}$ is the initial angular spread RMS, and $\sigma_{\Gamma}^{2}=\frac{\alpha}{\lambda^{2}}$
is the RMS of angular Gamma distribution. For the longitudinal energy RMS 
spread growth, we should also take the foil thickness variation into
consideration. Then the output energy RMS spread can be approximated
by Eq. \ref{LongiAngSp},
\begin{equation}
\sigma_{h}=\sqrt{\sigma_{f}^{2}+E_{1}^{2}+\sigma_{t}^{2}}\label{LongiAngSp}
\end{equation},
where $\sigma_{f}$ is the initial energy spread RMS, $E_{1}$ is the
energy distribution width calculated above, and $\sigma_{t}$ is the energy
spread RMS variation introduced by foil thickness variation. $\sigma_{t}$
can be calculated as $\sigma_{t}=\sigma_{w}t_{foil}E_{0,w}$. 
$\sigma_{w}$ is the RMS foil variation in percentage, $t_{foil}$ is the
foil thickness, and $E_{0,w}$ is the first derivation of $E_{0}$ over
the foil thickness. $E_{0,w}$ can be approximated by $\frac{E_{02}-E_{01}}{t_{2}-t_{1}}$,
where $E_{02}$ and $E_{01}$ are fit results of $E_{0}$ over two
different SRIM simulation with two different stripping foil thickness
$t_{2}$ and $t_{1}$.

\subsection{Misalignment}

Mis-alignments of beam elements are modeled by the Euclidian group
for 3D translations
\begin{equation}
\bar{x}\rightarrow\bar{x}-\Delta\bar{x}
\end{equation}
 and rotations
\begin{equation}
R_{\mathrm{3D}}\left(\bar{\theta}\right)=\left[\begin{array}{ccc}
m_{11} & m_{12} & m_{13}\\
m_{21} & m_{22} & m_{23}\\
m_{31} & m_{32} & m_{33}
\end{array}\right]=R_{z}\left(\theta_{z}\right)R_{y}\left(\theta_{y}\right)R_{x}\left(\theta_{x}\right)
\end{equation}
where
\begin{align*}
R_{x}\left(\theta_{x}\right) & =\left[\begin{array}{ccc}
1 & 0 & 0\\
0 & \cos\left(\theta_{x}\right) & \sin\left(\theta_{x}\right)\\
0 & -\sin\left(\theta_{x}\right) & \cos\left(\theta_{x}\right)
\end{array}\right],\\
R_{y}\left(\theta_{y}\right) & =\left[\begin{array}{ccc}
\cos\left(\theta_{y}\right) & 0 & \sin\left(\theta_{y}\right)\\
0 & 1 & 0\\
-\sin\left(\theta_{y}\right) & 0 & \cos\left(\theta_{y}\right)
\end{array}\right],\\
R_{z}\left(\theta_{z}\right) & =\left[\begin{array}{ccc}
\cos\left(\theta_{z}\right) & \sin\left(\theta_{z}\right) & 0\\
-\sin\left(\theta_{z}\right) & \cos\left(\theta_{z}\right) & 0\\
0 & 0 & 1
\end{array}\right]
\end{align*}
and
\begin{align*}
m_{11} & =\cos\left(-\theta_{y}\right)\cos\left(\theta_{z}\right),\\
m_{12} & =\sin\left(\theta_{x}\right)\sin\left(-\theta_{y}\right)\cos\left(\theta_{z}\right)+\cos\left(\theta_{x}\right)\sin\left(\theta_{z}\right),\\
m_{13} & =-\cos\left(\theta_{x}\right)\sin\left(-\theta_{y}\right)\cos\left(\theta_{z}\right)+\sin\left(\theta_{x}\right)\sin\left(\theta_{z}\right),\\
m_{21} & =-\cos\left(-\theta_{y}\right)\sin\left(\theta_{z}\right),\\
m_{22} & =-\sin\left(\theta_{x}\right)\sin\left(-\theta_{y}\right)\sin\left(\theta_{z}\right)+\cos\left(\theta_{x}\right)\cos\left(\theta_{z}\right),\\
m_{23} & =\cos\left(\theta_{x}\right)\sin\left(-\theta_{y}\right)\sin\left(\theta_{z}\right)+\sin\left(\theta_{x}\right)\cos\left(\theta_{z}\right),\\
m_{31} & =\sin\left(-\theta_{y}\right),\\
m_{32} & =-\sin\left(\theta_{x}\right)\cos\left(-\theta_{y}\right),\\
m_{33} & =\cos\left(\theta_{x}\right)\cos\left(-\theta_{y}\right)
\end{align*},
with Cartesian rotation angles: $\bar{\theta}\equiv\left[\mathrm{tilt,\mathrm{yaw},\mathrm{roll}}\right]$.

The matrix for 6D phase-space translations is
\begin{equation}
T_{\mathrm{ps}}\left(\Delta\bar{x}\right)=\left[\begin{array}{ccccccc}
1 & 0 & 0 & 0 & 0 & 0 & -\Delta x\\
0 & 1 & 0 & 0 & 0 & 0 & 0\\
0 & 0 & 1 & 0 & 0 & 0 & -\Delta y\\
0 & 0 & 0 & 1 & 0 & 0 & 0\\
0 & 0 & 0 & 0 & 1 & 0 & -k_{0}\Delta z\\
0 & 0 & 0 & 0 & 0 & 1 & 0\\
0 & 0 & 0 & 0 & 0 & 0 & 1
\end{array}\right]
\end{equation}
and for rotations,
\begin{equation}
R_{\mathrm{ps}}\left(\bar{\theta}\right)=\left[\begin{array}{ccccccc}
m_{11} & 0 & m_{12} & 0 & m_{13} & 0 & 0\\
0 & m_{11} & 0 & m_{12} & 0 & m_{13} & 0\\
m_{21} & 0 & m_{22} & 0 & m_{23} & 0 & 0\\
0 & m_{21} & 0 & m_{22} & 0 & m_{23} & 0\\
m_{31} & 0 & m_{32} & 0 & m_{33} & 0 & 0\\
0 & m_{31} & 0 & m_{32} & 0 & m_{33} & 0\\
0 & 0 & 0 & 0 & 0 & 0 & 1
\end{array}\right]
\end{equation}.
The new transport matrix is
\begin{align}
M\rightarrow & T_{\mathrm{ps}}^{-1}\left(\nicefrac{L}{2}\right)T_{\mathrm{ps}}^{-1}\left(\Delta\bar{x}\right)R_{\mathrm{ps}}^{-1}\left(\bar{\theta}\right)M\nonumber \\
 & \times R_{\mathrm{ps}}\left(\bar{\theta}\right)T_{\mathrm{ps}}\left(\Delta\bar{x}\right)T_{\mathrm{ps}}\left(\nicefrac{L}{2}\right)\\
= & \left(R_{\mathrm{ps}}\left(\bar{\theta}\right)T_{\mathrm{ps}}\left(\Delta\bar{x}\right)T_{\mathrm{ps}}\left(\nicefrac{L}{2}\right)\right)^{-1}M\\
 & \times R_{\mathrm{ps}}\left(\bar{\theta}\right)T_{\mathrm{ps}}\left(\Delta\bar{x}\right)T_{\mathrm{ps}}\left(\nicefrac{L}{2}\right)
\end{align}
where $T_{\mathrm{ps}}\left(\nicefrac{L}{2}\right)$ is a translation
to the element center.

\subsection{Multi-Charge-State Acceleration}

FRIB is accelerating multiple charge states simultaneously in order
to enhance the beam current. Accelerations of multiple charge states can be
modeled by FLAME via a divide-and-conquer scheme. A three-step scheme
is proposed:

Step 1 -- Machine initialization by reference charge state: An ideal
particle with a central charge state is used to initialize the whole
machine, mainly the cavity phase. The result is then recorded as the global
beam reference.

Step 2 -- Multi-charge-state beam reference orbit initialization: Single
particle tracking is used for a reference particle for each charge
state beam center and the result is recorded as the local beam reference
for each charge state.

Step 3 -- Envelope tracking for each charge state: The beam envelope
is then tracked using a transport matrix, which is adjusted according
to different charge states and reference orbits.

Sometimes the information for the beam center and the beam RMS envelope of the whole
beam are required. Then the beam recombination scheme is needed to
provide this information. The total beam center can be calculated
as Eq. \ref{MBcen},
\begin{equation}
\bar{x}=\frac{\sum_{i}N_{i}\bar{x}_{i}}{N}\label{MBcen}
\end{equation},
where $\bar{x}$ is the beam center of the whole beam and $\bar{x}_{i}$
is the beam center for each charge state beam. $N$ is the total number
of particles and $N_{i}$ is the number of particles for a certain
charge state. The total beam RMS can be calculated as Eq. \ref{MBrms},
\begin{equation}
\sigma=\sqrt{\frac{\sum_{i}N_{i}(\sigma_{i}^{2}+\bar{x}_{i}^{2})}{N}}\label{MBrms}
\end{equation},
where $\sigma$ is the whole beam RMS envelope size, and $\sigma_{i}$ is the RMS
beam envelope size for each charge state. The method and the benchmark have been
thoroughly discussed in~\cite{he2014linear}.

\section{Software Design of FLAME}

The idea of building a light-weighted online model which can handle
FRIB special physics challenges dates back to 2012~\cite{he2012analytical}.
The first working version aiming at testing the physics modeling concept
was written in Matlab. Then, the second version of the online model
prototype aiming at initial concept verification of online model
support for beam commissioning and physics application
started its development at the end of 2014~\cite{shen2015physics}.
The second prototype was called Thin Lens Model (TLM) ~\cite{shen2015development},
and was written in Java, with a python interface linking to the physics
application prototyping environment with Jpype~\cite{Jpype}. After concept verification of
TLM, we started development of FLAME at the
end of 2015, which is written in C++ with control room quality. In
this section, the software design strategy of FLAME is described.

\begin{figure}
\includegraphics[width=3.6in,height=0.8in]{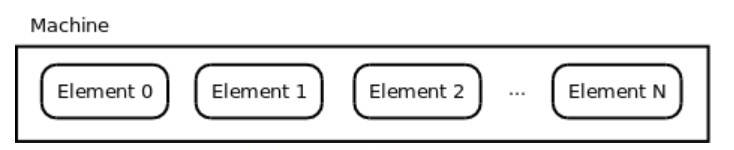} \caption{Elements in a Machine}

\label{Machine} 
\includegraphics[width=3.6in,height=1.1in]{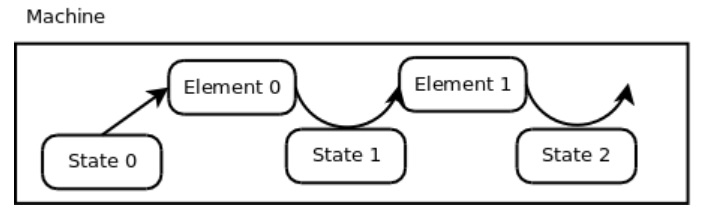} \caption{Propagation of State through a list of Elements}

\label{State} 
\end{figure}

FLAME contains a C++ core for efficient numerical calculation and
a Python wrapper for convenient work flow control. The core
concepts of the simulation engine are the Machine, Element, State,
 and Config. A Config is a container for key/value pairs. This is
the interface through which Elements are specialized. The lattice
file parser populates a Config, which may then be used to construct
a Machine. A Machine represents an ordered list of Elements (Fig. \ref{Machine})
~\cite{FlameDoc}.

A State represents ``the beam'', a particle or bunch of particles
which pass from Element to Element through a Machine. It keeps track
of important beam parameters such as the charge state, the accumulated phase,
the kinetic energy, the beam orbit vector and the beam envelope matrix. The operation
of the simulation is for an Element to transform a State. This new
(output) state is then passed as an input to the next element (Fig. \ref{State}).

FLAME uses a free style input lattice file which contains one or more
variable definitions, element definitions and beam line definitions.
Variables, for example -- charge states, initial kinetic energy  and initial
beam parameters -- can be defined as 
\[
<var\_name>=<value>;
\],
where \textless{}value\textgreater{} can be a string, a scalar, vector
floating point values or expressions of these types. An element definition
takes the form
\[
<instance>:<element\_type>,<key>=<value>,...;
\],
supported element type by now including: ``drift'', ``marker'', ``sbend'' (for the sector bend), ``quadrupole'',
``solenoid'', ``rfcavity'' (for the RF cavity), ``stripper'', ``edipole'' (for the electrostatic dipole) and ``equad'' (for the electrostatic quadrupole).
A beam line definition is constructed from zero or more elements, or
beam line names: 
\[
<instance>:LINE=(<element>,...);
\]
After that, a ``USE'' statement may be given to select which beam
line is expanded and used. 
\[
USE:<instance>;
\]

FLAME also has a high level python wrapper for flexible and easy control
of FLAME work flow. FLAME python interface allows ``on-the-fly'' read
and write access of the beam and lattice parameter without reloading most
data which is already in the memory. And the python interface also
makes many online beam tuning tasks, such as, running arbitrary segments
of lattice, lattice parameter optimization and development
of a virtual accelerator, which is convenient and efficient for developers. FLAME
also has a matrix caching capability. If FLAME infers no change of a transfer
matrix between two runs, it'll reuse the cached matrix instead of
recalculating it, which would also contribute to speeding up the calculation
process.

\section{Precision and Speed Benchmark of FLAME}

In this section, the calculation precision and calculation speed are
benchmarked with FRIB design codes. 

In our case, the lattice design codes, which are DIMAD~\cite{servranckx1984dimad} for the
front-end segment and IMPACT-Z for the main linac segment, are used to verify the simulation
precision. The calculation speed of FLAME is then discussed and benchmarked
with IMPACT-Z. Note that because the length of the front-end segment is short,
there is no issue for calculation speed no matter which code we choose, so we only focus on the linac segment which is designed using IMPACT-Z when discussing speed benchmark.

\subsection{Precision}

\begin{figure}
\centering \includegraphics[width=3.4in,height=1.5in]{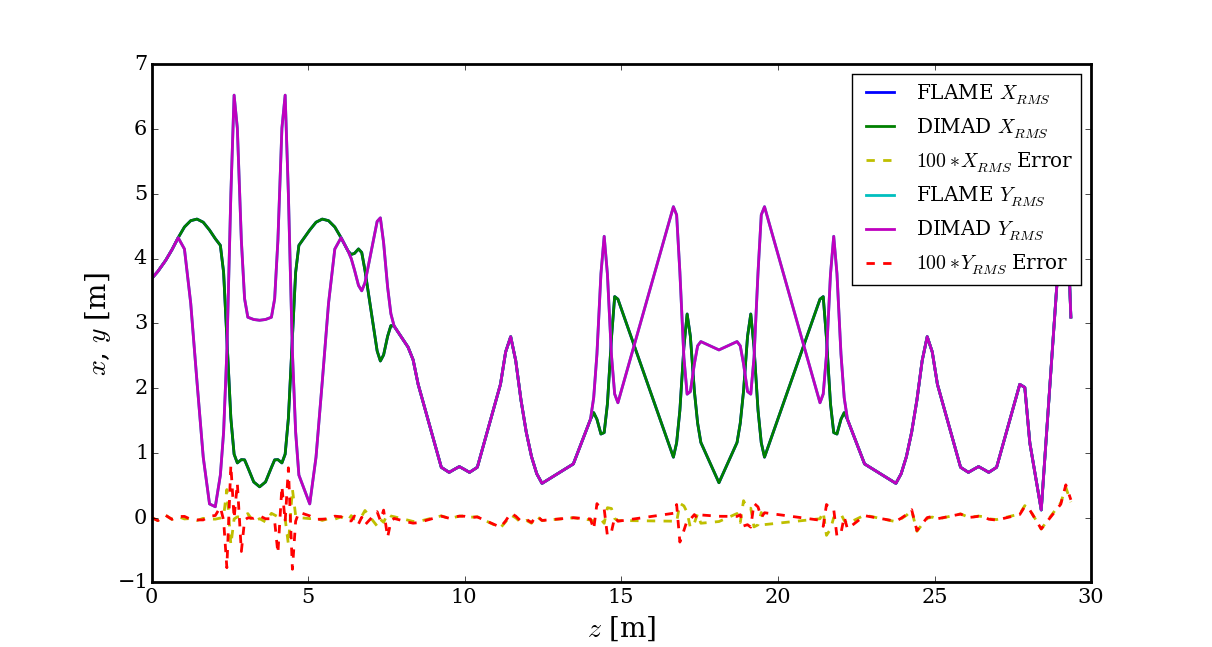}
\caption{The benchmark result between FLAME and DIMAD for the transverse beta-function
of front-end segment; blue: FLAME horizontal case; green: DIMAD horizontal case;
dashed yellow: 100 times magnified difference between FLAME and DIMAD
for horizontal beta-function; cyan: FLAME vertical case; magenta:
DIMAD vertical case; dashed red: 100 times magnified error between
FLAME and IMPACT-Z for vertical beta-function.}

\label{FlameBench1} 
\end{figure}

For the front-end segment, the results are benchmarked against DIMAD code.
The results for both horizontal and vertical beta-functions are shown
in Fig. \ref{FlameBench1}. The RMS error is $1.18\times10^{-3}$ m
for the horizontal beta-function and $1.97\times10^{-3}$ m for the vertical
beta-function. Because the FRIB front-end segment is manipulating coasting
beam before the radio-frequency quadrupole, calculation of the longitudinal beam profile is not needed.

\begin{figure}
\centering \subfigure{ \label{FlameBencha} 
\includegraphics[width=3.4in,height=1.5in]{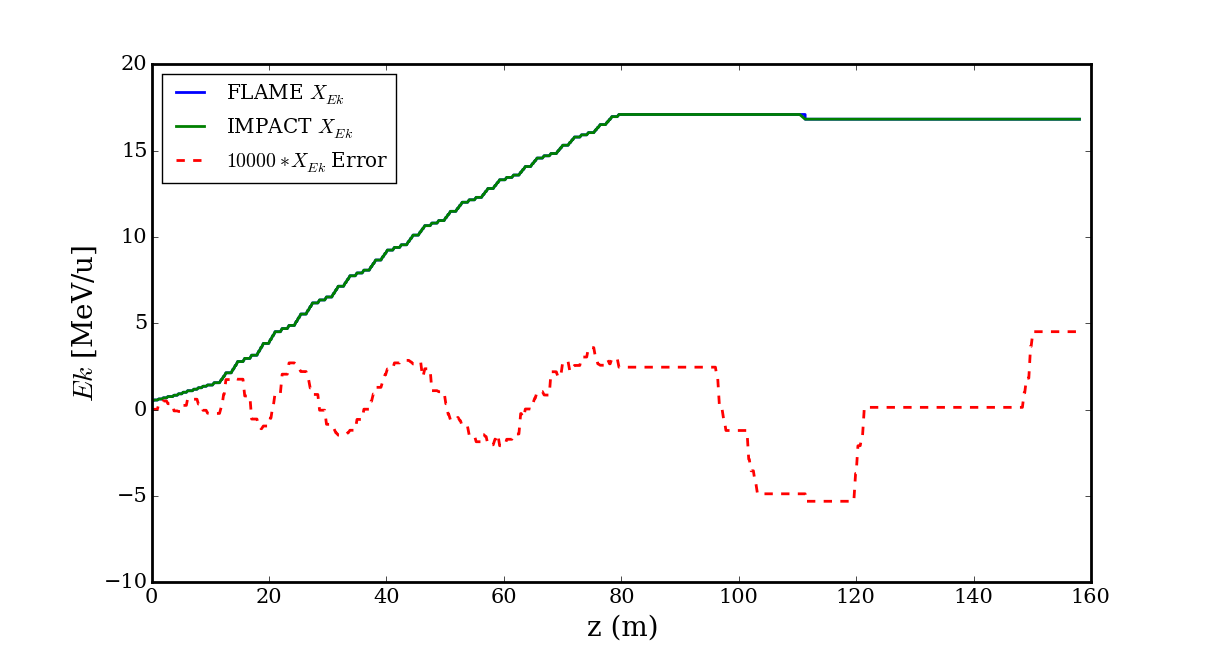} } \subfigure{
\label{FlameBenchb} 
\includegraphics[width=3.4in,height=1.5in]{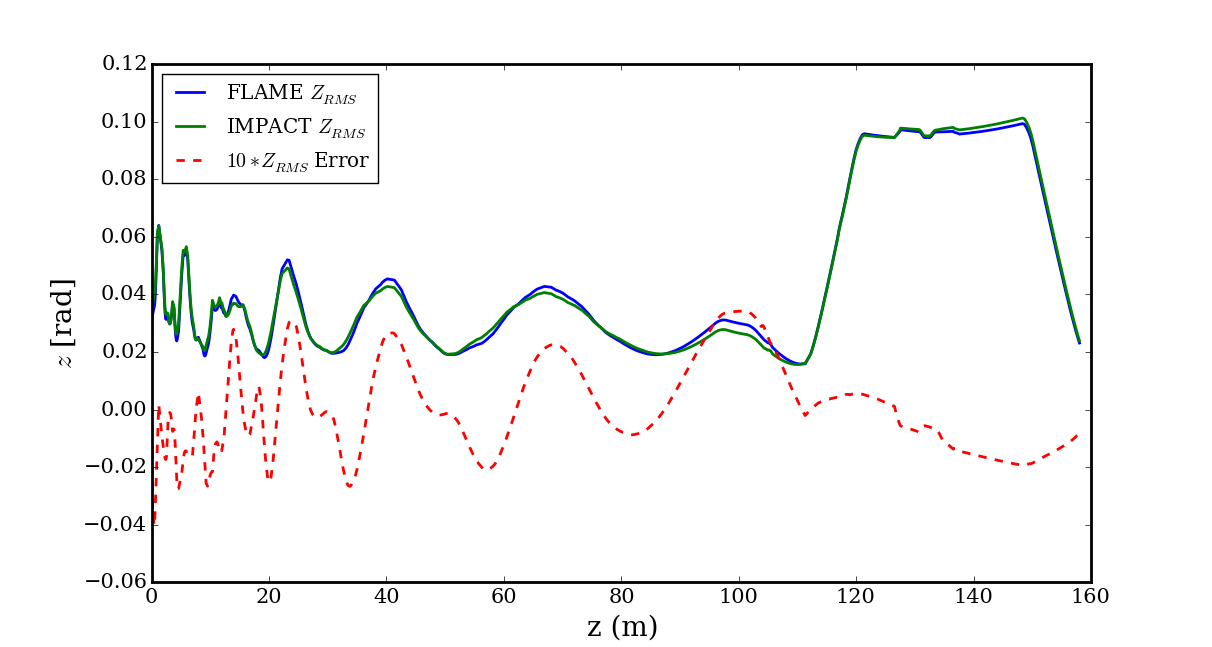} }
\subfigure{ \label{FlameBenchc} 
\includegraphics[width=3.4in,height=1.5in]{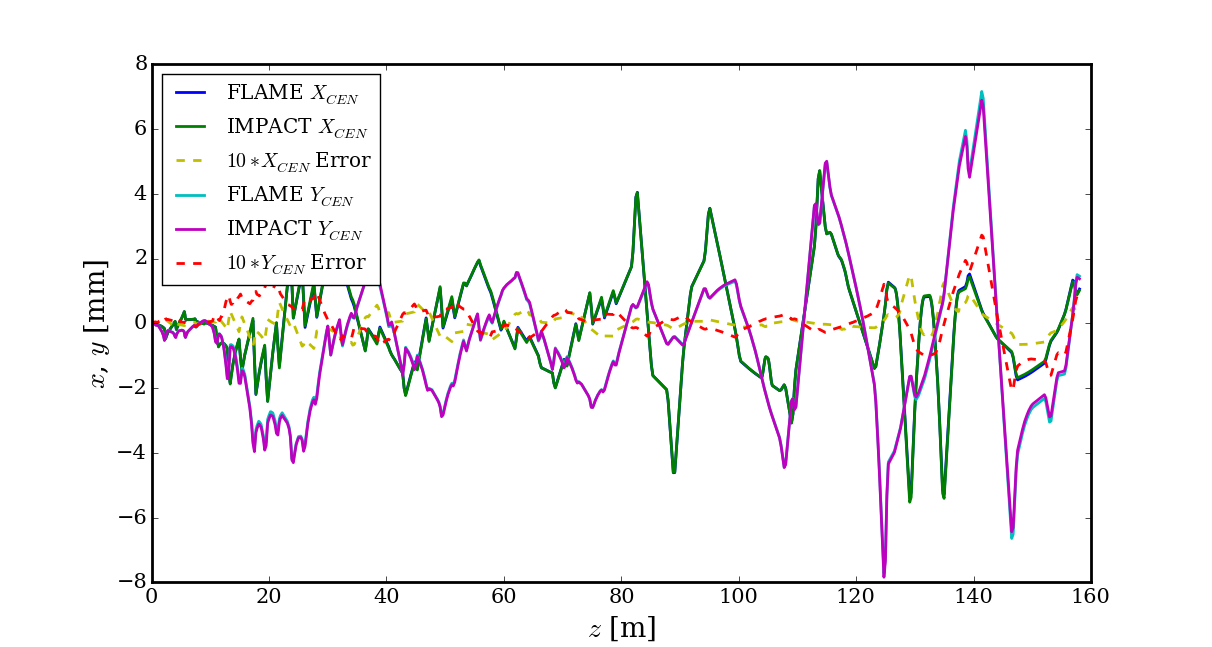} }
\subfigure{ \label{FlameBenchd} 
\includegraphics[width=3.4in,height=1.5in]{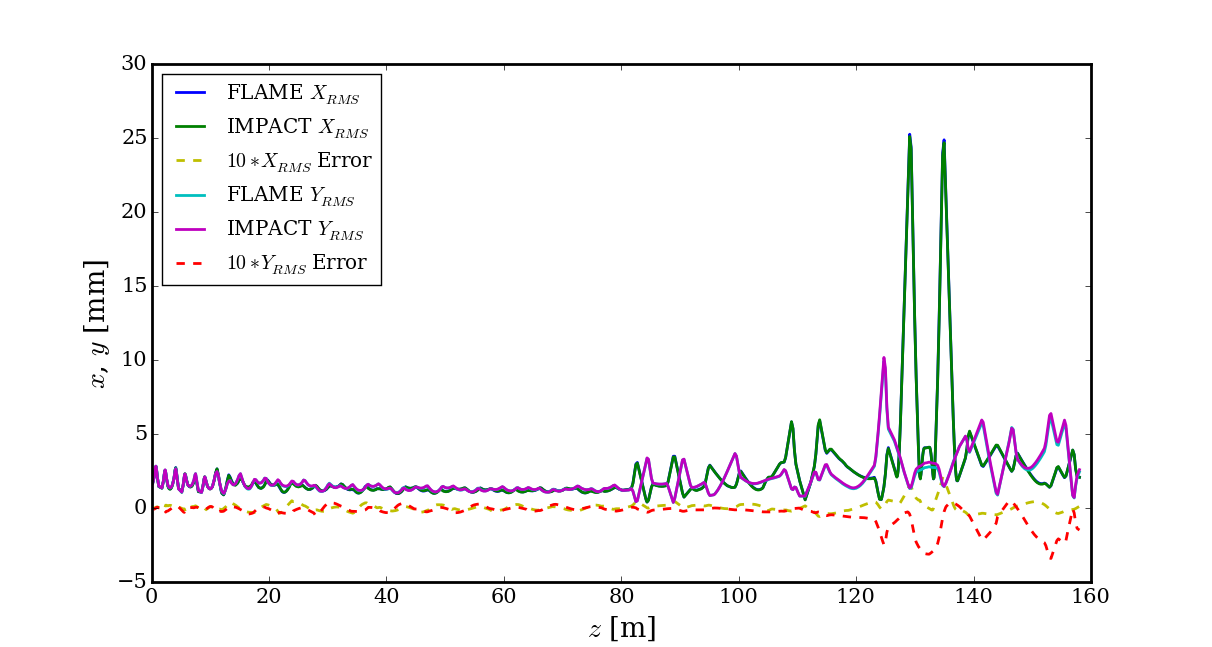} } \caption{Benchmark result between FLAME and IMPACT-Z for LS1+FS1. (a) The benchmark
result for the kinetic energy; blue: FLAME result, green: IMPACT-Z result;
dashed red: 10000 times magnified difference. (b) The benchmark result
for the longitudinal RMS size; blue: FLAME result; green: IMPACT-Z result,
dashed red: 10 times magnified difference. (c) The benchmark result for
the central beam orbit; blue: FLAME horizontal case; green: IMPACT-Z horizontal
case; dashed yellow: 10 times magnified difference between FLAME and
IMPACT-Z for the horizontal beam orbit; cyan : FLAME vertical case; magenta:
IMPACT-Z vertical case; dashed red: 10 times magnified difference between
FLAME and IMPACT-Z for the vertical beam orbit. (d) The benchmark result for
transverse beam RMS size; blue: FLAME horizontal case; green: IMPACT-Z
horizontal case; dashed yellow: 10 times magnified difference between
FLAME and IMPACT-Z for the horizontal beam size; cyan: FLAME vertical case;
magenta: IMPACT-Z vertical case; dashed red: 10 times magnified error
between FLAME and IMPACT-Z for the vertical beam size.}

\label{FlameBench2} 
\end{figure}

For the main linac, the result needs to be benchmarked in both transverse
and longitudinal directions against IMPACT-Z code. The linac segment
1 (LS1) plus folding segment 1 (FS1) of FRIB lattice are used. Uranium
33+ and 34+ before the charge stripper and 76+ to 80+ after the charge stripper
are accelerated, and the beam center and the beam envelope after multi-charge-state
recombination are benchmarked. The 3D field tracking method is used to handle 
RF cavities in IMPACT-Z. Fig. \ref{FlameBencha} shows the benchmark
result for the kinetic energy, $2.54\times10^{-4}$ MeV/u RMS error can
be achieved. Fig. \ref{FlameBenchb} shows the benchmark result for the 
longitudinal RMS size. The RMS error is $1.55\times10^{-3}$ rad. Fig.
\ref{FlameBenchc} shows the benchmark result for the beam orbit with cavity
dipole components considered. The RMS error is 0.107 mm for the horizontal
direction and 0.128 mm for the vertical direction. Fig. \ref{FlameBenchd}
shows the benchmark result for the transverse beam RMS size with cavity quadrupole
components considered. The RMS error is 0.091 mm for the horizontal
direction and 0.075 mm for the vertical direction.

\subsection{Speed}

One important design goal for an online model is to ensure fast
speed while keeping reasonably high precision. Table \ref{Speed}
shows FLAME running time under various circumstances. A personal laptop with
Intel(R) Core(TM) i7-6820HQ CPU @ 2.70 GHz (4 cores 8 threads) is
used. FLAME is able to retain all lattice and cavity data information
inside the memory, which is easy to edit and reuse. Therefore the initialization
time is excluded from the total running time. From column 1 and column
2, we can see that, both C++ API and Python API of FLAME is able to
finish calculation within several tens of milli-seconds. And matrix
caching is an efficient way of boosting speed of FLAME between two
runs where element matrices do not change.

Because IMPACT-Z is a particle tracking code aiming at off-line accelerator
design, simply running it will not give satisfactory speed. For example,
if we use 20K particles, 3D field tracking based cavity model with
integration step 60 to simulate LS1+FS1, the total time it takes is
27 s. However, comparing this time with FLAME does not make sense
because this time is not the pure calculation time, and one can always
decrease the number of particles and integration steps to increase the speed
at the expense of precision. To make a fair comparison between IMPACT-Z
and FLAME, we take both the speed and precision into consideration, and
make the following assumptions to IMPACT-Z: further improvement to
IMPACT-Z can be easily made so that (1) IMPACT-Z can keep the large
3D E\&M file in the memory for reuse, so the 3D file loading time (around
12s) can be subtracted from the total time; (2) IMPACT-Z will no longer
output the data onto the disk and the file I/O time can be subtracted from
the total time; (3) IMPACT-Z will reuse the lattice file which is kept
in the memory, so the initialization time can be subtracted from the total
time. For FLAME, similar assumptions have already been realized, so
no extra assumptions are needed.

\begin{table*}[b]
\caption{\label{Speed}Running time of FLAME under various circumstances }

\begin{ruledtabular} %
\begin{tabular}{lccc}
Performance Benchmark  & %
\begin{tabular}{c}
C++ API {[}msec{]}\tabularnewline
\end{tabular} & %
\begin{tabular}{c}
Python API {[}msec{]}\tabularnewline
\end{tabular} & %
\begin{tabular}{c}
Matrix Cached Run {[}msec{]}\tabularnewline
\end{tabular}\tabularnewline
\colrule 1 charge state for LS1  & 16.0  & 17.7  & 6.4\tabularnewline
2 charge states for LS1  & 26.3  & 28.3  & 6.5\tabularnewline
1 to 5 charge states for LS1+FS1  & 25.4  & 28.2  & 8.0\tabularnewline
2 to 5 charge states for LS1+FS1  & 36.0  & 39.2  & 8.2\tabularnewline
\end{tabular}\end{ruledtabular} 
\end{table*}

\begin{figure}
\begin{minipage}[t]{0.5\linewidth}%
 \centering \subfigure{ \label{Fig6a2} 
\includegraphics[width=1.7in,height=1.5in]{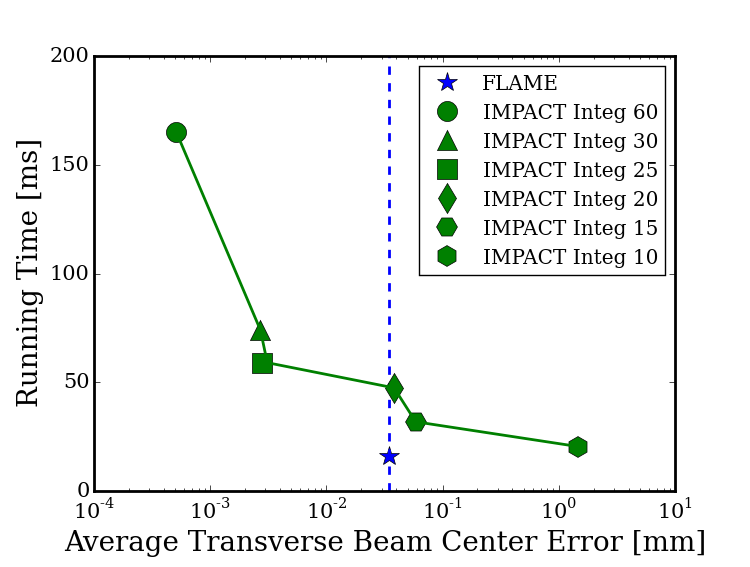}
} %
\end{minipage}%
\begin{minipage}[t]{0.5\linewidth}%
 \subfigure{ \label{Fig6b2} 
\includegraphics[width=1.7in,height=1.5in]{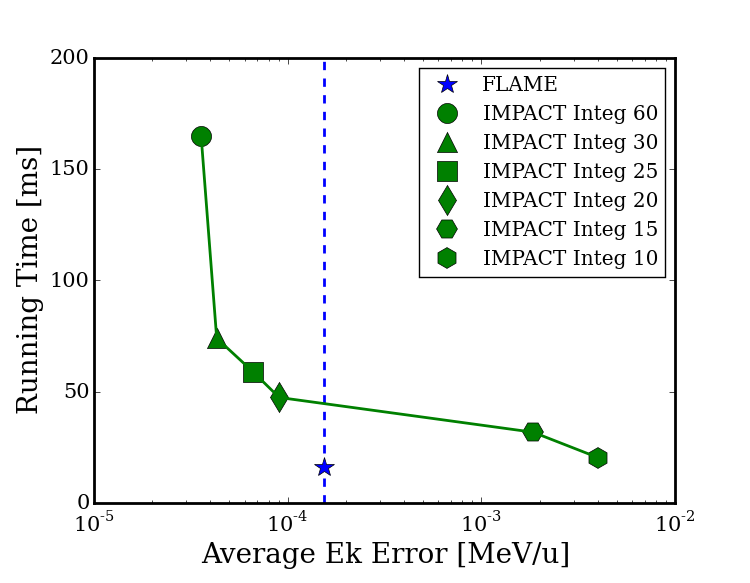}
} %
\end{minipage}\caption{Comparison of the calculation precision and time consumption of beam orbit
and kinetic energy between FLAME and IMPACT-Z for a single particle; blue:
FLAME case; green: IMPACT-Z case with various integration steps for
RF cavity.}

\label{SpeedBenchOrbit} 

\begin{minipage}[t]{0.5\linewidth}%
\centering \subfigure{ 
\includegraphics[width=1.7in,height=1.5in]{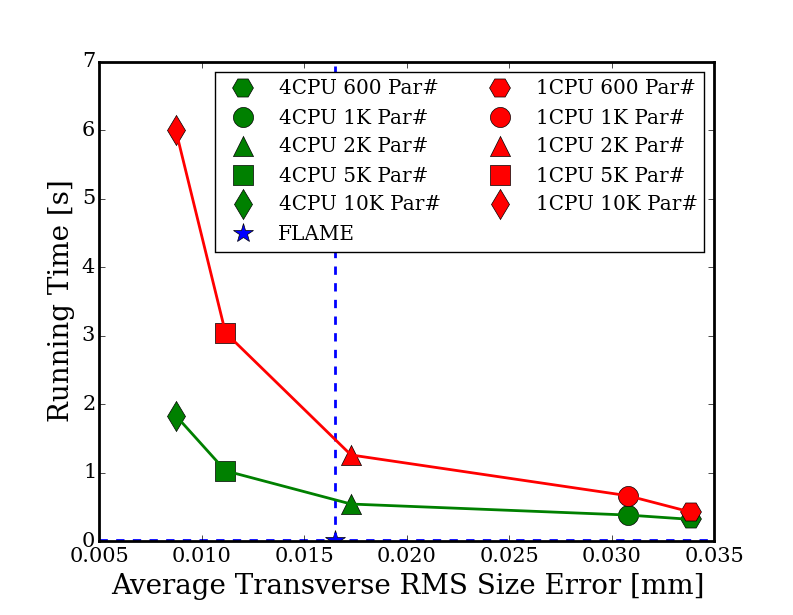}
} %
\end{minipage}%
\begin{minipage}[t]{0.5\linewidth}%
\subfigure{ 
\includegraphics[width=1.7in,height=1.5in]{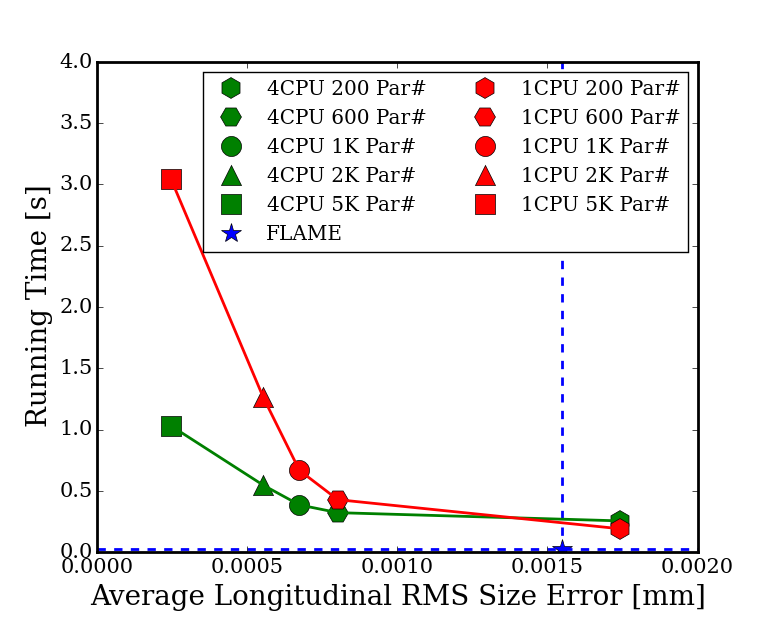}
} %
\end{minipage}\caption{Comparison of the calculation precision and the time consumption of transverse
and longitudinal beam RMS between FLAME and IMPACT-Z; blue: FLAME case; 
red: IMPACT-Z case with various particle number, 1 core case; green:
IMPACT-Z case with various particle number, 4 cores case.}

\label{SpeedBenchEnv} 
\end{figure}

\textbf{Beam Orbit and Energy}: IMPACT-Z can use a single particle to
calculate the beam orbit and the reference energy. The integration step of
all elements except the RF cavities are set to 1, and the integration step
of the RF cavity is treated as a variable. We set the integration step 1000
as a reference case, where the calculation results already converge. The lattice
of LS1 is used. Both FLAME and IMPACT-Z are run on the same laptop
computer described above. The calculation precision and speed are then
compared with FLAME (Fig. \ref{SpeedBenchOrbit}). Precision of FLAME
is comparable to IMPACT-Z with integration steps 20 while 3 times
faster. Note that FLAME already gets the envelope information within
this period of time as a bonus.

\textbf{Beam Envelope}: The precision of the envelope calculation of IMPACT-Z
depends on particle numbers. In this study, the particle number is
treated as a variable. We set particle number 100K as a reference
case where results already converge. The integration step of all elements
except RF cavities is set to 1 and the integration step of RF cavity
is set to 30. Lattice of LS1 with 2 charge states is used. Because
IMPACT-Z supports message passing interface (MPI) parallelization, we also tested single core and
4 cores cases. The calculation result and speed are then compared with
FLAME (Fig. \ref{SpeedBenchEnv}). According to the result, the precision
of FLAME is comparable to IMPACT-Z with 2K particles for the transverse
directions while 48 times faster than 1 core case and 21 times faster
than 4 cores case. For longitudinal direction, the precision of FLAME is comparable
to IMPACT-Z run with 200 particles and 7 times faster than 1 core case,
and 10 times faster than 4 cores case. (MPI overhead may drag down the overall efficiency of IMPACT-Z when the particle number is small)

\section{CONCLUSION}

A new software, FLAME, has been developed for the purpose of FRIB
online model service and physics application. FLAME is able to handle
the special physics challenges of modeling non-axisymmetric RF cavities
and multi-charge state acceleration encountered by FRIB. A multipole
expansion based thin-lens cavity model has been thoroughly discussed
to handle the dipole and quadrupole terms produced by the non-axisymmetric
RF cavities. FLAME contains a C++ core and a python interface for 
easy access and work flow control.
Calculation result benchmark and performance study confirm that FLAME
has achieved good balance among precision, speed and
convenience, and can be a strong tool in support of FRIB beam
commissioning tasks.
\begin{acknowledgments}
The authors would like to thank Yan Zhang at FRIB for initiative insight
and discussion on an online model specially designed for FRIB and
support on precursor prototypes development; Prof. Chuanxiang Tang
at Tsinghua University, on inspiring discussing on how to handle cavity
multipole components; J. Qiang, the author of the IMPACT-Z code, on
useful information about using IMPACT-Z and understanding its physics
behind; Z. Zheng at FRIB for providing 3D cavity field data. The author
would also like to thank Z. Liu, J. Wei, S. Lund, Q. Zhao, Y. Yamazaki
and F. Marti at FRIB for numerous helpful discussions,
and the MSU writing center for providing language support of the paper. 
The work is supported by the U.S. National Science Foundation under Grant No.
PHY-11-02511, and the U.S. Department of Energy Office of Science
under Cooperative Agreement DE-SC0000661. 
\end{acknowledgments}

\bibliographystyle{apsrev4-1}
\bibliography{PRSTAB_FLAME.bib}

\end{document}